\documentclass{aa}
\usepackage{ulem}
\usepackage{amsmath}
\usepackage{graphicx}
\usepackage{natbib}
\usepackage{multicol}
\usepackage{longtable}
\usepackage{graphicx}
\usepackage{supertabular,booktabs}
\usepackage{sidecap}
\usepackage{enumitem}
\usepackage{txfonts}
\usepackage{epstopdf}
\usepackage{tabularx}
\usepackage{longtable}
\usepackage{ltablex}
\usepackage[pdftex]{color}
\usepackage[]{natbib}

\DeclareUnicodeCharacter{2212}{}
\begin{document}
%\maketitle
\title{Sr and Ba abundance determinations: comparing machine-learning with star-by-star analyses}
\subtitle{High-resolution re-analysis  
of suspected  LAMOST barium stars}
%A word of caution about machine-learning studies using Sr lines

\author{D. Karinkuzhi
          \inst{1,2}
          \and
          S. Van Eck\inst{2}
          \and
          A. Jorissen\inst{2}
          \and
          A. Escorza\inst{3}
          \and
          S. Shetye\inst{2}
          \and
          T. Merle\inst{2}
          \and 
          L. Siess\inst{2}
          \and 
          S. Goriely\inst{2}
          \and
          H. Van Winckel\inst{4}
          }

 \institute{Department of Physics, Indian Institute of Science, Bangalore 560012, India
           \and
            Institut d'Astronomie et d'Astrophysique, Universit\'{e} Libre de Bruxelles C.P. 226, Boulevard du Triomphe, B-1050 Bruxelles, Belgium
              \and
      European Southern Observatory, Alonso de Córdova 3107, Vitacura, Casilla 19001, Santiago de Chile, Chile
\and
Instituut voor Sterrenkunde, KULeuven, Celestijnenlaan 200D, 3001 Leuven, Belgium
             }

   \date{Received X, 2021; accepted Y, 2021}
\abstract
{A new large sample of 895 s-process-rich candidates out of 454\;180 giant stars surveyed by LAMOST at low spectral resolution ($R \sim 1800$) has been reported by Norfolk et al. (2019; hereafter N19).
%Norfolk et al. (2019; hereafter N19). 
    % context (optional)
}
{We aim at confirming the s-process enrichment at the higher resolution ($R \sim 86\;000$) offered by the HERMES-Mercator spectrograph, for the 15 brightest targets of the 
%N19 
previous study sample which consists in 13 Sr-only stars and two Ba-only stars (this terminology designates stars with  either only Sr or only Ba lines strengthened).
    % aims
}
{Abundances were derived  for elements 
Li, C (including $^{12}$C/$^{13}$C isotopic ratio), N, O, Na, Mg, Fe, Rb, Sr, Y, Zr, Nb, Ba, La, and Ce, using the  TURBOSPECTRUM radiative transfer  LTE code   with  MARCS  model  atmospheres. Binarity has been tested by comparing the Gaia DR2 radial velocity (epoch 2015.5) with the HERMES velocity obtained 1600 - 1800~d (about 4.5 years) later. 
    % methods
}
{Among the 15 programme stars, four show no s-process overabundances ([X/Fe] $ < 0.2$~dex), eight show mild s-process overabundances (at least three heavy elements with $0.2 < \rm{[X/Fe]} < 0.8$), and three have strong overabundances (at least three heavy elements with [X/Fe] $\geq$ 0.8). Among the 13 stars classified as Sr-only by the previous investigation, four have no s-process overabundances, eight are mild barium stars, and one is a strong barium star. The two Ba-only stars turn out to be both strong barium stars. Especially noteworthy is the fact that these two are actually {\it dwarf} barium stars. Two among the three strong barium stars show clear evidence for being binaries, as expected for objects produced through mass-transfer. The results for the no s-process and mild barium stars  are more surprising. Among the no-s stars, there are two binaries out of four, whereas only one out of the eight diagnosed mild barium stars  show a clear signature of radial-velocity variations. 
}
{Blending effects and saturated lines have to be considered very carefully when using machine-learning techniques, especially on low-resolution spectra.
Among the Sr-only stars from the previous study sample, one may expect about 
60\% (8/13) of them to be true mild barium stars and about 8\% to be strong barium stars, and this fraction is likely close to 100\% for the previous study Ba-only stars (2/2).
It is therefore recommended to restrict to the previous study Ba-only stars when one needs an unpolluted sample of mass-transfer (i.e., extrinsic) objects.

}
    % conclusion (optional)


\maketitle

\section{Introduction}

Barium (Ba) stars or \ion{Ba}{II} stars, as they were originally named, are G- and K-type giants with strong absorption lines of slow-neutron-capture (s)-process elements in their spectra, in combination with enhanced carbon-bearing molecular 
bands. They were first identified as chemically peculiar by \citet{Bidelman1951}, who discussed their distinctive spectroscopic characteristics and stressed the extraordinary strength of the resonance line of ionised barium at 4554~\AA. The resulting overabundance of barium and other s-process elements on the surface of these stars could not be explained from an evolutionary point of view because the s-process of nucleosynthesis takes place in the interiors of Asymptotic Giant Branch (AGB) stars, whereas Ba stars are instead dwarf, subgiant, red-giant-branch (RGB), or red-clump stars \citep[e.g.,][]{Jorissen2019,Escorza2019}. Barium stars are  understood to originate from a binary evolution channel \citep{McClure1983}. According to this formation scenario, the carbon and the s-process elements were transferred to the current primary from a more evolved companion when the latter was in its AGB phase.

Galactic chemical evolution provides an alternative explanation for mild barium stars (with [Ba/Fe] $\sim 0.2 - 0.3$~dex), which represent the  upper [Ba/Fe] tail of the Galactic ([Ba/Fe],[Fe/H]) trend \citep[e.g.][]{Edvardsson1993, Tautvaisiene-2021}.

Since recently, the largest homogeneous sample of barium stars was collected in the course of the Michigan Spectral Sky Survey, with 205 new discovered barium stars
\citep{MacConnell1972}.
Mainly based on this sample, \citet{Lu1983} then built their catalogue with 221 entries, followed by an updated version with 389 stars \citep{Lu1991}. However, a substantial fraction of them are probably not barium stars \citep[especially those classified with a Ba index\footnote{The Ba index (spanning the range 1 -- 5, later extended to 0 -- 5) has been defined by \citet{Warner1965} based on a visual inspection of the strength of the \ion{Ba}{II} 4554~\AA~ line, the index 5 corresponding to the strongest line strength.} $\le 1$; e.g.,][]{Smiljanic2007}. 

More recently, large-field spectroscopic surveys like LAMOST \citep{Wu2011,Bai2016}, involving low-resolution spectroscopy, have permitted to potentially  increase in a tremendous manner the  number of known stars with enhanced s-process elements. For instance,  \citet[hereafter N19]{Norfolk2019}, 
have reported 859 candidates (out of 454\,180 giants studied) which were classified as either Sr-only, or Ba-only, or Ba- and Sr-strong. This classification was based on the comparison between the strengths of the most conspicuous \ion{Sr}{ii} (4077  and 4215 \AA~) and \ion{Ba}{ii} lines (4554, 4934, and 6496 \AA~) in template and target stars, using the machine-learning technique {\it "The Cannon"} \citep{Ness2015}. There are however several caveats (as we discuss below) with this approach, which call for an {\it a posteriori} verification of the s-process enhancement from high-resolution spectra. Only one star was subject to such a check by N19.

The purpose of this paper is to perform such a verification on a larger sample of 15 stars. 
 The motivation thereof is 
the very low resolution of LAMOST spectra ($R \sim 1800$) combined with the fact that the above-mentioned  lines of \ion{Sr}{ii} and \ion{Ba}{ii} are 
known 
to show a positive luminosity effect \citep[i.e., strengthening of the line due to low gravity rather than overabundance;][]{Gray2009}. Moreover, some of those lines are blended (e.g., \ion{Sr}
{II}
4215.5~\AA~ by CN lines) and are often saturated, in which case they become poor abundance diagnostics. 

Recently, the Sr abundance in Carbon-Enhanced Metal-Poor (CEMP) stars has gained a lot of attention since some studies \citep{Hansen2016,Hansen2019} found that the Sr/Ba ratio can be used to separate  CEMP stars into their sub-groups (CEMP-no, CEMP-s, and CEMP-rs) and to identify their progenitors since this ratio depends on the nucleosynthetic sites. 

Large field spectroscopic surveys have provided spectra for millions of stars and machine-learning techniques are widely used to measure  abundances of the elements.  Our current analysis aims at discussing the difficulties in measuring the abundances of Ba and Sr especially when using machine-learning techniques on low-resolution spectra. 

In this paper, 
we present a detailed high-resolution spectroscopic analysis of the brightest s-process-rich candidates of N19 in order (i) to 
check for possible misclassification as (mild) barium stars, 
(ii) to understand the origin of the variations in their individual elemental abundance pattern and thereby understand the origin of these peculiar abundances, and (iii) to evaluate the power of machine-learning techniques for abundance determination from low-resolution spectra.

This paper is organized as follows. Section~\ref{Sect:sample} describes the selection of the sample. Section~\ref{Sect:parameters} discusses the method used for deriving the atmospheric parameters. Section~\ref{Sect:abundances} presents the abundance analysis.
Section~\ref{Sect:classification} compares N19 %\citet{Norfolk2019} 
classification with ours, whereas Sect.~\ref{Sect:Individual} presents comments about individual stars. Section~\ref{Sect:originofpattern} discusses the possible origin of the peculiarities of the different identified classes, and finally Sect.~\ref{Sect:TheCannon} discusses the efficiency with which the machine-learning method {\it The Cannon} has been able to correctly flag s-process-enriched stars from low-resolution spectra. Conclusions are presented in Sect.~\ref{Sect:conclusions}.

\section{Sample selection}
\label{Sect:sample}

Our analysis focuses on the brightest among the stars from N19 tagged as 'Sr only' or 'Ba only' candidates (this terminology
designates stars with either only Sr or only Ba lines strengthened, respectively; see Sects.~\ref{Sect:light-s} and \ref{Sect:heavy-s}), visible from the {\it Roque de los Muchachos Observatory} in La Palma, Canary Islands (Spain). They are listed in Table~\ref{Tab:programme_stars} along with N19 %\citet{Norfolk2019} 
classification.  They were observed with the high-resolution HERMES spectrograph \citep{Raskin2011} mounted on the 1.2m Mercator telescope. The spectra covers the spectral range 3900~--~9000~\AA~ with a resolution of 86,000. The S/N ratio of the HERMES spectra around 5000~\AA~ is listed in Table~\ref{Tab:programme_stars}. 

\section{Derivation of atmospheric parameters}
\label{Sect:parameters}

The atmospheric parameters of the programme stars were derived following the same method as outlined by \citet{Karinkuzhi2018}. We used the {\sc BACCHUS} (Brussels Automatic Code for Characterizing High accUracy Spectra) tool in a semi-automated mode \citep{Masseron2016}. {\sc BACCHUS} combines interpolated MARCS model atmospheres \citep{Gustafsson2008} with the 1D local-thermodynamical-equilibrium (LTE) spectrum-synthesis code {\sc TURBOSPECTRUM} \citep{Alvarez1998,Plez2012}. We manually selected \ion{Fe}{I} and \ion{Fe}{II} lines so as to choose
blending-free lines for {\sc BACCHUS}  to derive the stellar parameters ($T_{\rm eff}$, [Fe/H], $\log g$, microturbulence velocity $\xi$ as well as rotational velocity).
 The code includes on the fly spectrum synthesis, local continuum normalization, estimation of local S/N ratio and automatic line masking.
 It computes
abundances using equivalent widths or spectral synthesis, allowing to check for excitation and ionization equilibria, thereby constraining $T_{\rm eff}$ and $\log g$. The microturbulent velocity $\xi$ is calculated by ensuring consistency between Fe abundances derived from lines of various reduced equivalent widths.
 \begin{table*}
\caption{Programme stars and adopted atmospheric parameters. 
}
\label{Tab:programme_stars}
\begin{tabular}{lllrcccccc}
\hline
\\
Name &  \multicolumn{1}{c}{$T_{\rm eff}$}&\multicolumn{1}{c}{$\log g$}  & \multicolumn{1}{c}{[Fe/H]}& $\xi$ &S/N& \multicolumn{1}{c}{Class}&  \\
     &\multicolumn{1}{c}{(K)}       &\multicolumn{1}{c}{(cm s$^{-2}$)}  &\multicolumn{1}{c}{(dex)}  &  (km s$^{-1}$) &&\multicolumn{1}{c}{}  &\\
\hline\\
\multicolumn{8}{c}{\bf no s-process enrichment}\medskip\\
HD 7863&4637 $\pm$ 64 & 2.29 $\pm$ 0.40& $-$0.07 $\pm$ 0.05&1.26 $\pm$ 0.10&68 &  no&& \\
&4561 $\pm$ 6	 &	2.37 $\pm$ 0.01	&	0.13 $\pm$ 0.01 &2& --  &  Sr only& \medskip\\
HIP 69788&5127 $\pm$ 11& 3.90 $\pm$ 0.14& $-$0.04 $\pm$ 0.04&0.61 $\pm$ 0.10 & 75& no && \\
& 4913 $\pm$ 	10 & 3.04 $\pm$ 0.02 & $-0.34 \pm			0.01$ & 2&--  &  Sr-only& \medskip\\
TYC 3144$-$1906$-$1&4136 $\pm$ 64 & 1.89 $\pm$ 0.50& $-$0.13 $\pm$ 0.10 & 1.37 $\pm$ 0.04& 48  &no (Li)&&  \\
& 4232 $\pm$ 8 & 1.87 $\pm$  0.02 & 0.15 $\pm$ 0.01 & 2&--&Sr only&&  \medskip\\
TYC 4684$-$2242$-$1&4651 $\pm$ 20 & 2.70 $\pm$ 0.14 & $-$0.05 $\pm$ 0.07& 1.15 $\pm$ 0.05&54&no&& \\
&4652 $\pm$ 12 & 2.71 $\pm$ 	0.03 & 0.05 $\pm$0.02 &2& --&Sr only&& 
\medskip\\
\multicolumn{8}{c}{\bf mild s-process enrichment}\medskip\\
BD $-07^\circ$ 402 &4654 $\pm$ 6 & 2.62 $\pm$ 0.19& $-$0.11 $\pm$ 0.05&1.22 $\pm$ 0.10 &61& mild (Li-rich)&& \\
& 4688  $\pm$ 9 &	2.58 $\pm$  0.02 &	0.11 $\pm$ 0.01 & 2&-- & Sr only& \medskip\\
BD $+44^\circ$ 575 & 4175 $\pm$ 6 & 1.50 $\pm$ 0.19& $-$0.45 $\pm$ 0.05& 1.60 $\pm$ 0.10 & 76&mild&&\\
&4202 $\pm$ 12 &	1.59  $\pm$ 0.03 &	$-0.11 \pm 0.02$ & 2&--& Sr only&& \medskip\\
TYC 22$-$155$-$1&4704 $\pm$ 9& 3.10 $\pm$ 0.32 &$-$0.20 $\pm$ 0.10 & 1.04 $\pm$ 0.05&47& mild&& \\
& 4629 $\pm$  11 &	2.72 $\pm$  0.03	& $-0.22 \pm  0.01$ & 2&--&Sr only& & \medskip\\
TYC 2913$-$1375$-$1&4757 $\pm$ 69& 2.00 $\pm$ 0.30 & $-$0.61 $\pm$ 0.11 & 1.45 $\pm$ 0.05&32&mild&&  \\
&4791$\pm$ 15 &	2.41$\pm$ 0.04 &	$-0.37 \pm0.02$ & 2&--&Sr only&  \medskip\\
TYC 3305$-$571$-$1&4816 $\pm$ 3 & 2.76 $\pm$ 0.16 & $-$0.05 $\pm$ 0.08 & 1.31 $\pm$ 0.04&49&mild&& \\
& 4798 $\pm$ 8 &	2.62 $\pm$ 0.02 & 0.18 $\pm$ 0.01	& 2&--&Sr only & \medskip\\
TYC 752$-$1944$-$1&5069 $\pm$ 25   & 2.94 $\pm$ 0.05 & $-$0.08 $\pm$ 0.08 &1.33 $\pm$ 0.04&61& mild&&\\
& 4967  $\pm$  11 &	2.79  $\pm$ 0.03 &	0.02  $\pm$ 	0.01 &2& -- & Sr only &
\medskip\\
TYC 4837$-$925$-$1&4679 $\pm$ 34   & 2.16 $\pm$ 0.29 & $-$0.27 $\pm$ 0.07& 1.30 $\pm$ 0.04&44&mild&&\\
&4739 $\pm$ 14 &2.46 $\pm$ 0.04 &	0.02 $\pm$ 	0.02 & 2&-- &Sr only&\medskip\\
TYC 3423$-$696$-$1&5042 $\pm$ 64 & 3.66 $\pm$ 0.30 & 0.02 $\pm$ 0.08 & 0.96 $\pm$ 0.04&55&mild && \\
&5014 $\pm$ 17&	3.59 $\pm$ 0.03&	0.22 $\pm$ 0.02&2&-- &Sr only && \medskip\\
\multicolumn{8}{c}{\bf strong s-process enrichment}\medskip\\
TYC 2250$-$1047$-$1&5335 $\pm$ 25& 3.71 $\pm$ 0.18 & $-$0.55 $\pm$ 0.12 & 1.45 $\pm$ 0.05&32&strong&& \\
& 5097 $\pm$ 23 & 	3.25  $\pm$ 0.03 &	$-0.68\pm0.03$ &2& -- &Ba only
& \medskip\\
TYC 2955$-$408$-$1&4716 $\pm$ 64 & 2.49 $\pm$ 0.3 & $-$0.39 $\pm$ 0.08 & 1.25 $\pm$ 0.04&61&strong&&  \\
& 4724 $\pm$ 10 & 	2.39  $\pm$ 0.03 &	$-0.21\pm 0.01$ & 2& --&Sr only&  \medskip\\
TYC 591$-$1090$-$1&5267 $\pm$ 36   & 3.68 $\pm$ 0.50 & $-$0.30 $\pm$ 0.12& 1.18 $\pm$ 0.06&28&strong&& \\
&5106 $\pm$ 13& 	3.33 $\pm$ 0.02 &	$-0.29\pm0.01$ & 2& -- &Ba only& \\
\hline
\end{tabular}
\tablefoot{$\xi$ is the microturbulence velocity. 
The column "S/N" gives the signal-to-noise ratio (around 5000~\AA~) of the HERMES spectrum used for the abundance analysis. The column "class" indicates classification either from our study (first line), or from N19 on the second line. For the criteria used to classify stars as 'no s-process', 'mild s-process enrichment' and 'strong s-process enrichment', see Sect.~\ref{Sect:classification}.}
\end{table*}

\setlength{\tabcolsep}{6pt}

\section{Abundance analysis}
\label{Sect:abundances}

Abundances are derived by comparing observed and synthetic spectra generated
with the TURBOSPECTRUM code.
The solar abundances are taken from \citet{Asplund2009}. We used the line lists assembled in the framework of the Gaia-ESO survey \citep{Heiter2015,Heiter2020}. These lines are presented in \citet{Karinkuzhi2018, Karinkuzhi2021}, hence we do not list them again here. The abundances are derived under the LTE assumption, but {\it a posteriori} NLTE corrections have been added whenever available, as we discuss below. In Table~\ref{Tab:light_abundances} and \ref{Tab:abundances}, we present all the abundances derived from our target stars. In the following, we comment on individual elemental abundances.

\subsection{Li}

The Li abundance has been derived from the \ion{Li}{I} 6707~\AA~ line. We could measure the Li abundance in only two stars, TYC 3144$-$1906$-$1 and BD $-07^\circ$402 with $\log \epsilon({\rm Li}) \approx 0.6$ and 1.3~dex respectively (Table~\ref{Tab:abundances}). These values are in accordance with the Li abundance of 1.0~dex predicted in RGB after the first dredge up
\citep[e.g.,][and references therein]{Jorissen2020}

\subsection{{\it C, N}, and {\it O}}
\label{Sect:CNO}

 We derive oxygen abundances 
from the [\ion{O}{I}] line at 6300.303~\AA~ 
except for TYC~591$-$1090$-$1 where the \ion{O}{I} resonance triplet at 7774~\AA~  is used instead. 
A non-LTE correction of 0.2 dex has been applied to obtain the final adopted O abundance for this object \citep{Asplund2005,Amarsi2016}. In TYC~2913$-$1375$-$1 and TYC~3144$-$1906$-$1, we could detect neither the  6300.303~\AA~ line nor the 7774~\AA~ line. Hence we used another $\alpha$-element, namely Ca, and adopted [Ca/Fe] as a proxy for [O/Fe] (Table ~\ref{Tab:light_abundances}).

The carbon abundance  is  obtained mainly from the CH band at 4310~\AA~ and from the C$_2$ bands at 5165 and 5635~\AA. Since our programme stars do not show strong enrichment of carbon, the C$_2$ bands  are not saturated.  We could derive consistent abundances from these three bands.

The nitrogen abundance for the programme stars are derived from the CN bands above 7500~\AA. 
The $^{12}$C/$^{13}$C ratio is derived using  $^{12}$CN features at 8003.553 and
 8003.910~\AA~, and $^{13}$CN features at 8004.554, 8004.728, 8004.781, 8010.458, and 8016.429~\AA. For several stars, the signal-to-noise ratio was not high enough to enable us to estimate the $^{12}$C/$^{13}$C ratio.

\subsection{Light s-process elements: Sr, Y and Zr}
\label{Sect:light-s}

The Y abundances for the programme stars are determined from the \ion{Y}{II} lines. 
The Zr abundance is derived using \ion{Zr}{I} and \ion{Zr}{II} lines, which yield consistent abundances. 

We now present a detailed discussion of all the lines involved in the Sr abundance determination, either by us or by N19, as Sr is a key element in N19 barium-star diagnostic.
In the present work, the Sr abundance is estimated using the \ion{Sr}{I} lines at 4607.327~\AA~  
(resonance line), 4811.877~\AA~   (non-resonant) and 7070.070~\AA~ (non resonant).  

For the \ion{Sr}{I} line at 4811.877 \AA~ \citep[not used by][]{Karinkuzhi2018, Karinkuzhi2021}, a $\log gf$ of 0.190 has been used \citep{Garcia1988}. For the 4607.327 \ion{Sr}{I} line, \citet{Hansen2013} advocate the value of 0.283 for its $\log gf$ \citep[from][]{Parkinson1976}, because it allows to match the solar Sr abundance.
An analysis of the HERMES Arcturus spectrum shows a similar agreement as for the Sun, as revealed by the first line of Table~\ref{Tab:Sr}. Adopting a metallicity of $-0.62$ for Arcturus \citep{Maeckle1975}, the 4607 and 4812 lines 
yield [Sr/Fe]~=~$-0.43$ and $-0.20$, respectively, in agreement with    
\citet{Maeckle1975} who found  [Sr/Fe] $ = -0.4\pm0.3$~dex. 

The \ion{Sr}{I} line at 4607.327~\AA~is known to form under NLTE conditions \citep{Bergemann2012,Hansen2013}, and the former authors  list in their Table~3 the NLTE corrections for this line at  metallicities 0.0 and $-$0.60 for various temperatures and surface gravities. The atmospheric parameters of all our programme stars are 
within this range, and the  
corresponding NLTE corrections  vary between 0.1 and 0.27 dex.  In Table~\ref{Tab:Sr}, we list separately  the
abundances derived from the three
clean (i.e., unblended and not saturated) Sr lines, namely \ion{Sr}{I} 4607.327, 4811.877~\AA~   and
7070.070~\AA, along with the NLTE correction (between parentheses) applied to the LTE abundance from the 4607 line (for the latter line, Table~\ref{Tab:Sr} lists the NLTE abundance).
 Table~\ref{Tab:abundances} provides the average [Sr/Fe] abundance as derived from these three lines.

\begin{figure}
\includegraphics[width=9cm]{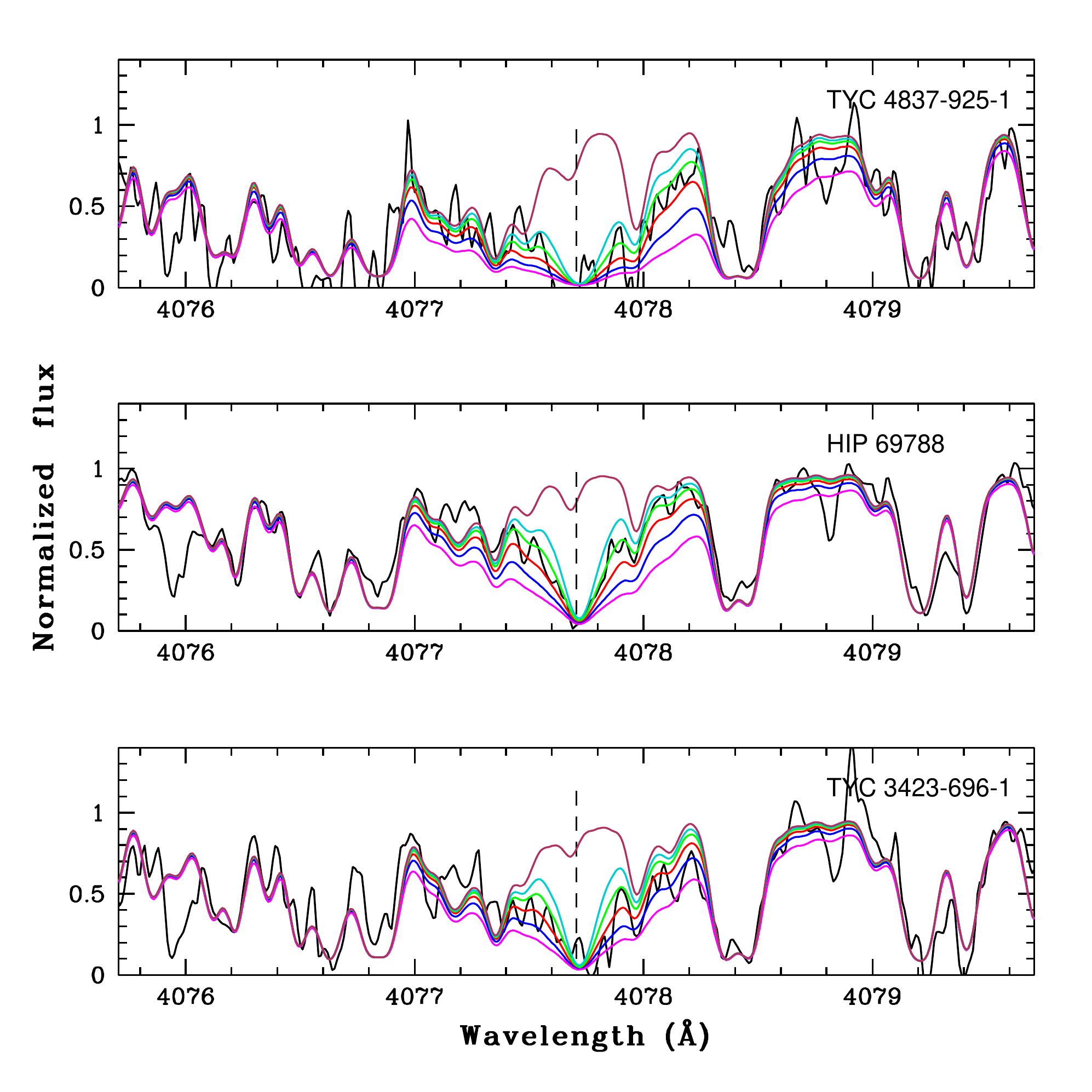}
\includegraphics[width=9cm]{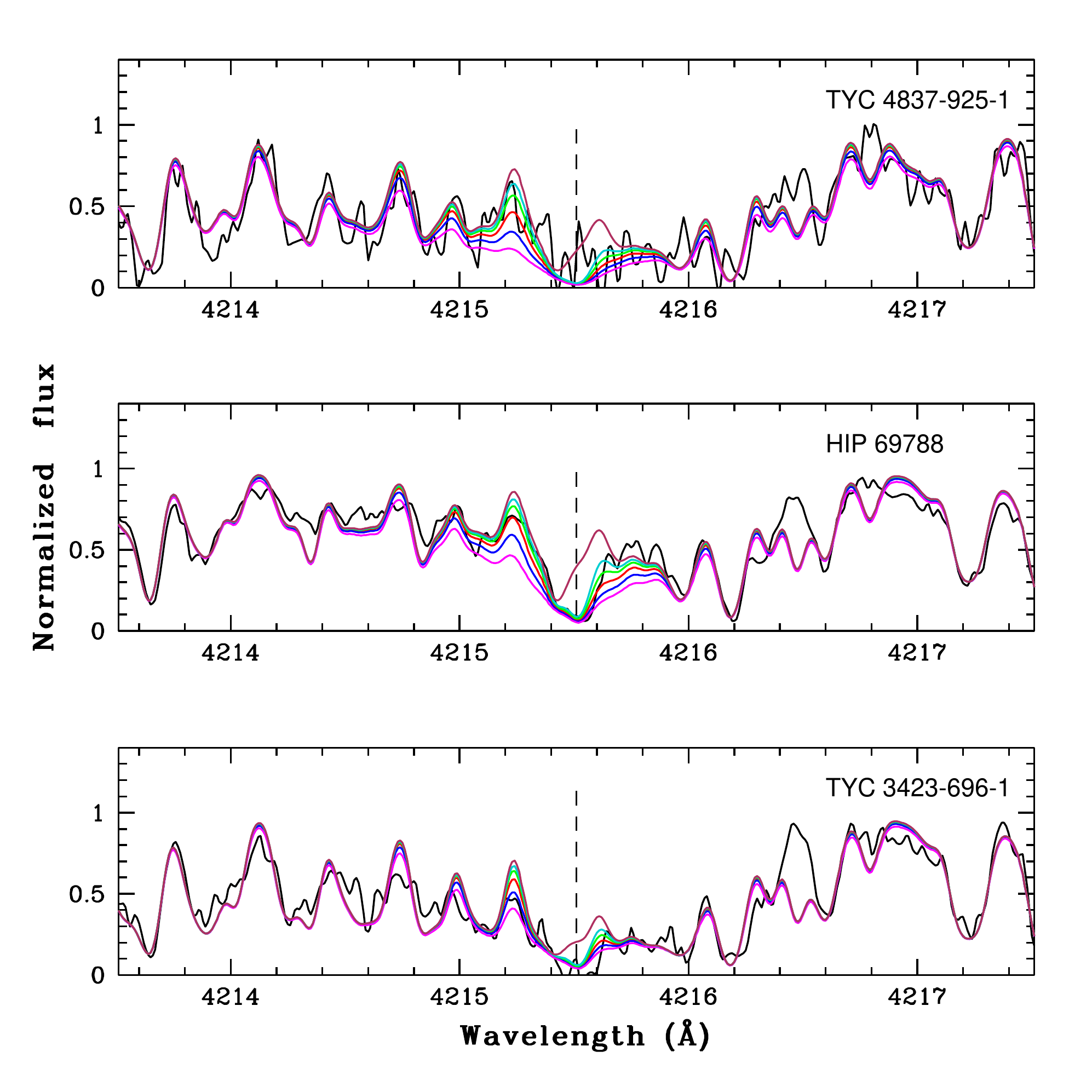}
\caption{The \ion{Sr}{II} 4077.7~\AA~ (top panel) and \ion{Sr}{II} 4215.5~\AA~ (bottom panel) lines  are shown in three N19 "Sr-only" stars. The magenta, blue, red, green and turquoise lines correspond to $\log \epsilon$(Sr) = 3.6, 3.3, 3.0, 2.7 and 2.4 dex, respectively. The brown line is for no Sr. These lines are clearly saturated. 
}
\label{Fig:Sr-line}
\end{figure}

To derive the Sr abundances, %\citet{Norfolk2019} 
N19 used instead the \ion{Sr}{II} lines at 
4077.077 and
4215.519~\AA. In stars of solar and mildly subsolar metallicities, as it is the case for all the programme stars, these lines are however saturated  \citep[Fig.~\ref{Fig:Sr-line}; see also][]{Hansen2013}, and could not be used to derive abundances.

\begin{figure}
\includegraphics[width=9cm]{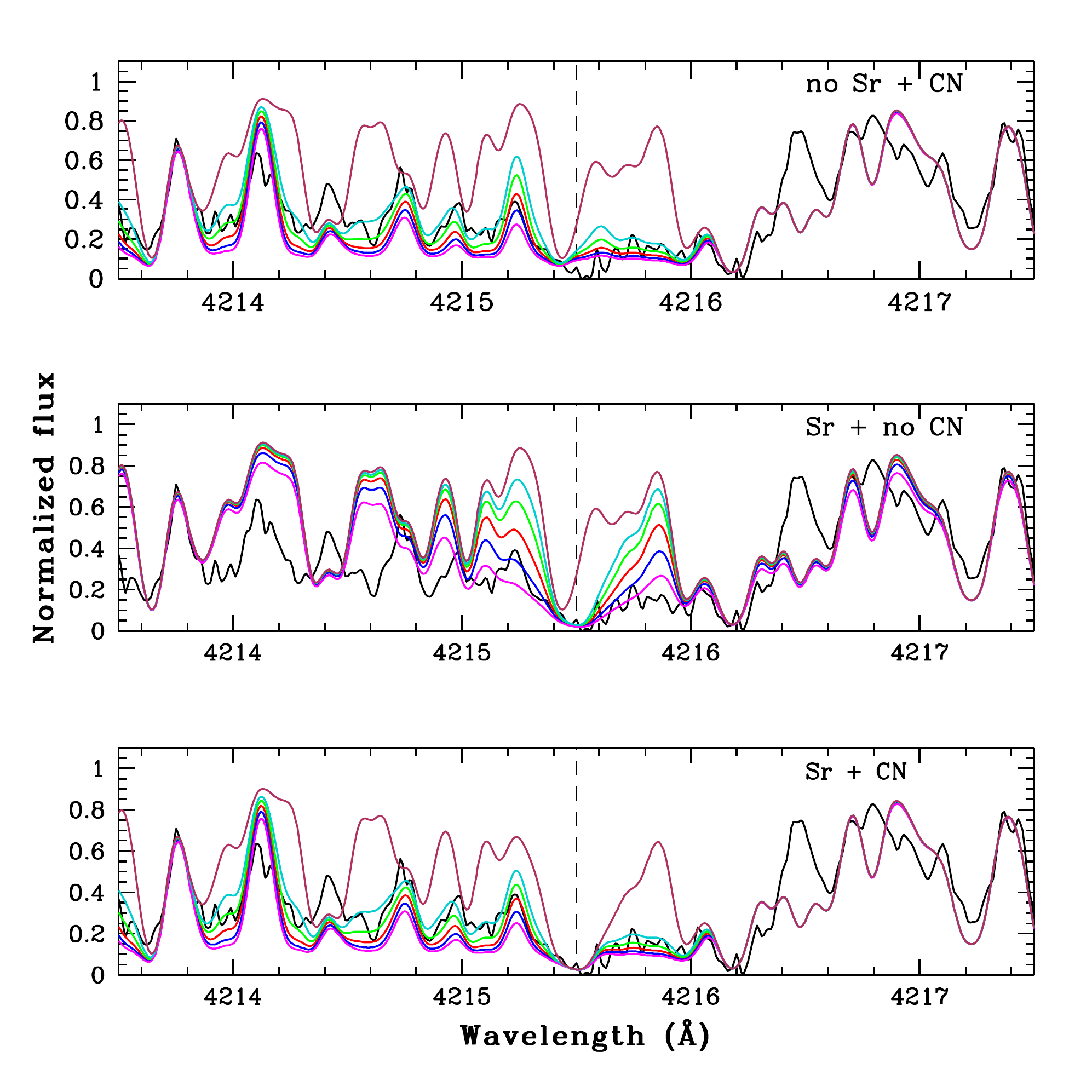}
\caption{Spectral syntheses of the 4214 -- 4217~\AA~  wavelength range surrounding the \ion{Sr}{II} 4215.5~\AA~  line in HD~7863.
In the top panel, the spectral syntheses do not include the \ion{Sr}{II} 4215.5~\AA~ line, to reveal the impact of the 4216~\AA~CN band head. The magenta, blue, red,
green  and turquoise  lines correspond  to $\log\epsilon$
(N)~
=~9.15, 8.85, 8.55, 8.25
and 7.95 dex, respectively. In the middle panel, the contribution from the CN band is removed and spectral syntheses are done by varying only the Sr abundance. In this case, the magenta, blue, red,
green  and turquoise  lines correspond  to $\log \epsilon$
(Sr)~
=~3.6, 3.3, 3.0, 2.7
and 2.4 dex, respectively. In the bottom panel, the synthesis includes both Sr and CN, with the adopted N abundance of $\log \epsilon$(N) = 8.55. The color-coding of the Sr abundance is the same as in the middle panel.  The adopted $^{12}$C/$^{13}$C ratio is 19.}
\label{Fig:Sr-noCN}
\end{figure}

Nevertheless, for the sake of comparison, Table~\ref{Tab:Sr} lists the Sr abundances provided by N19
for these lines, along with our very uncertain abundance estimate from the 4077.077~\AA~ line. The  4215.519~\AA~ line could not be used to derive even a rough abundance estimate as the spectrum syntheses for the different abundances lie on top of each other (Fig.~\ref{Fig:Sr-line}). This situation is due to the presence of the strong CN bandhead at $\lambda$ 4216~\AA~ \citep[Sect.~27 of][]{Gray2009}, which strongly depresses the continuum, especially in stars with enhanced C or N. Figure~\ref{Fig:Sr-noCN} shows the strong impact of the CN band in the 4215~\AA~  region, even in the absence of the Sr line. Note especially how the "no Sr + CN" (top panel) and "Sr + CN" (bottom panel) are barely distinguishable, even at the high resolution of HERMES, which is nearly 50 times the one of LAMOST.

\begin{figure}
\includegraphics[width=9cm]{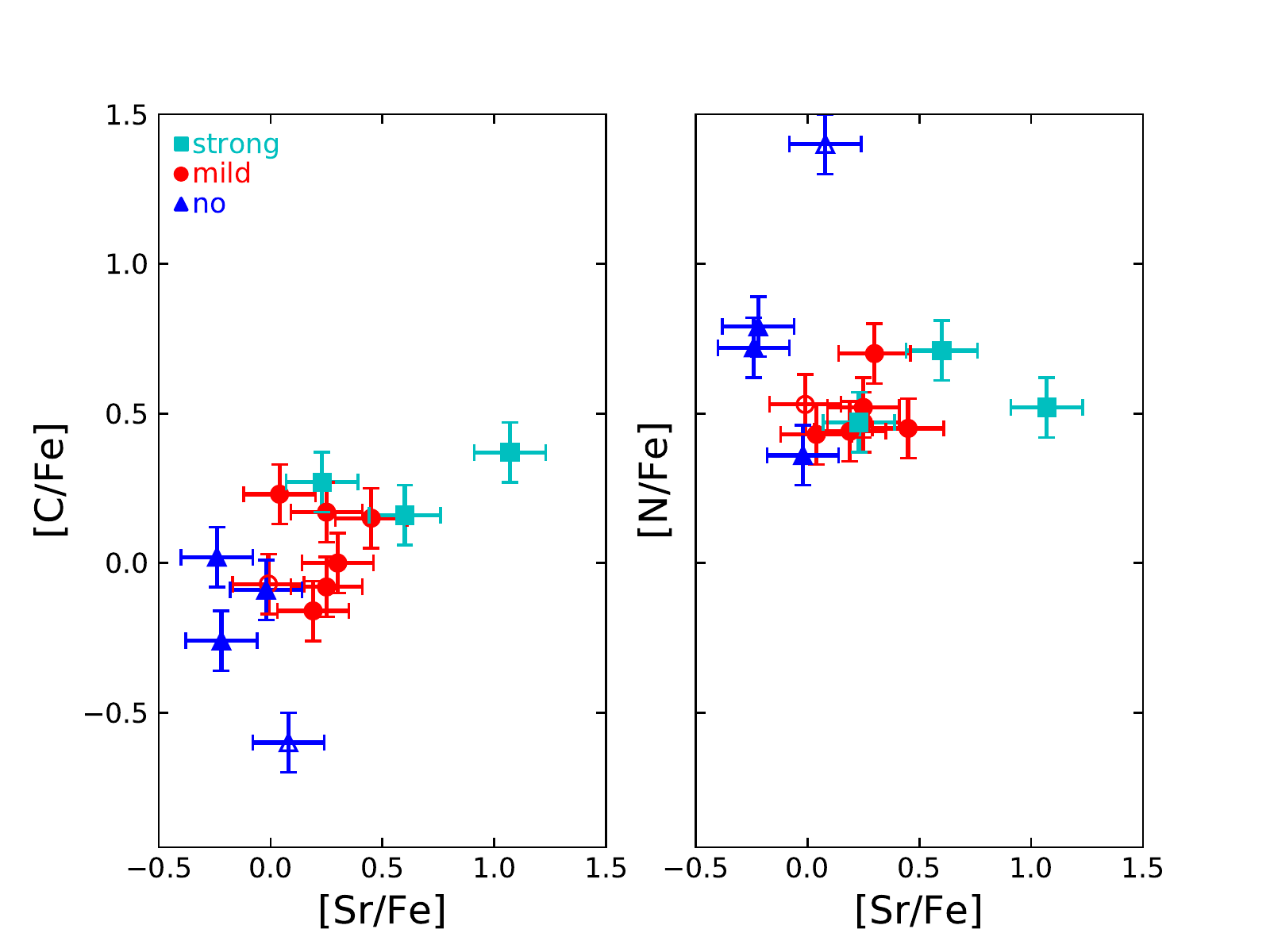}
\caption{The correlation between the [Sr/Fe] index and the C and N abundances, showing a clear N overabundance for three among the four Sr-no stars (see Sect.~\ref{Sect:classification}.}
\label{Fig:Sr-N}
\end{figure}

In this respect, it is certainly relevant to note that the three no-s stars,  considered as Sr-only by N19  (TYC~3144$-$1906$-$1, TYC~4684$-$2242$-$1, and HD~7863; see Table~\ref{Tab:programme_stars} and Sect.~\ref{Sect:classification}), are precisely those being N-rich (Fig.~\ref{Fig:Sr-N}).

\begin{table*}
\caption[]{Line by line abundances of Sr and Ba in the programme stars.
%, in units of $\log\epsilon$(H) = 12.
}
\small{
\begin{tabular}{lrrrrrrrrrrrrrr}
\hline\\
                      & \multicolumn{3}{c}{[\ion{Sr}{I}/H]}&& \multicolumn{2}{c}{[\ion{Sr}{II}/H]}&& \multicolumn{5}{c}{[\ion{Ba}{II}/H]} \\
\cline{2-4}\cline{6-7}\cline{9-13}\\
     \multicolumn{1}{r}{$\lambda$(\AA~)}                 &   4607.33 & 4811.88 &  7070.07&&  4077.08 &  4215.52  &&   4524.92 &  4554.03 & 4934.08 & 5853.67 &  6141.67\\
\cline{2-4}\cline{6-7}\cline{9-13}\\
Arcturus & $-$1.05 (0.27) & $-$0.82 \medskip\\
&\multicolumn{10}{c}{\bf no s-process enrichment}\medskip\\
HD 7863&  $-$0.27 (0.2) &  $-$0.27 & $-$0.27    && $-$0.17: &-- &&  0.12& $-$0.18: & -- &  $-$0.18& $-$0.18 \\
       &  --   &--  & --     && 0.9  & 0.9      &&--      & 0.0   &0.1   &--      &--
\medskip\\
HIP 69788 & $-$0.22 (0.1)  &0.13 &--      &&$-$0.17:&-- &&--       &0.12:&0.12:&0.12       &--  \\
          &-- &--    &--     &&0.7   &0.5       &&--      &0.7   &0.4 &--      &--\medskip\\
TYC 3144$-$1906$-$1   &-- &  $-$0.12 &$-$0.02    &&$-$0.17:  &--      &&--&0.12: &0.12: &$-$0.18      &$-$0.18\\
                  &--   &  &--     && 1.0   & 1.0     &&--      &0.2   &0.0&--      &--\medskip\\
TYC 4684$-$2242$-$1 &$-$0.57 (0.2)&-- &$-$0.02    &&0.13:  &--        &&-- &0.12: &0.12: &0.02      &0.02\\
            &-- &--    &--     &&0.9   &0.9 &&--      &-0.1   &0.0   &--      &--\medskip\\
&\multicolumn{10}{c}{\bf mild s-process enrichment}\medskip\\
BD $-07^\circ$402  &$-$0.27 (0.2)&0.12 & $-$0.17     && 0.13:&--  &&-- & 0.12:& 0.22:&0.1 &0.1\\
      &--  &--   &--     &&0.9  &1.0   &&--      &0.0&-0.1&--      &--\medskip\\
BD +44$^\circ$575 &--&$-$0.2   &$-$0.17    &&$-$0.17:      &--         &&--      &0.12:&0.12:&$-$0.2&--\\
      &--  &--   &--     &&0.9&0.8    &&--      &$-$0.2&$-$0.1&--      &--\medskip\\
TYC 22$-$155$-$1 &$-$0.37 (0.2) &$-$0.2 &0.13     &&$-$0.17:      &--         &&0.3: &0.12:&0.12:&0.1&0.1\\
      &--  &--   &--     &&0.5&0.5    &&--      &-0.2&-0.3&--      &--\medskip\\
      &--  &--   &--     &&0.2&0.4  &&--      &$-$0.3&$-$0.5&--      &--\medskip\\
TYC 2913$-$1375$-$1 &\tablefootmark{a}-- & --&--     && --      &--      &&--&$-$0.18&$-$0.03:&$-$0.68      &$-$0.68\\
      &--  &--   &--     &&0.2&0.4  &&--      &$-$0.3&$-$0.5&--      &--\medskip\\
TYC 3305$-$571$-$1  &0.30 (0.2)&--& 0.13:     && -- &--      &&--      &0.42:&--&0.32      &0.32\\
      & -- &--    & --     &&0.9&0.9 &&--      &0.2&0.1&--      &--\medskip\\
TYC 752$-$1944$-$1 &0.03 (0.1) &--&0.43     &&--      &--         &&--&0.62:&0.62:&0.52      &0.52\\
      & --     & -- &--    &&0.8&0.8 &&--      &0.3&0.0&--      &--\medskip\\
TYC 4837$-$925$-$1  &$-$0.27(0.2) &-- &0.13 &&$-$0.17: &--         &&--  &0.12:&0.12:&$-$0.18&$-$0.18\\
      &--   &--  &--     &&0.8&0.8  &&--      &0.0&-0.1&--      &--\medskip\\
TYC 3423$-$696$-$1 &0.53: (0.1)&0.43&--     &&$-$0.17:      &--          &&-- &0.12:&0.12:&0.0 &0.1\\
      &--    &-- &--     &&0.8   &0.9 &&--      &0.1&0.2&--      &--\medskip\\
&\multicolumn{10}{c}{\bf strong s-process enrichment}\medskip\\
TYC 2250$-$1047$-$1 &0.33 (0.2) &0.73&--     &&--      &--      &&0.62 &0.82:&0.82:&0.62      &0.62\\
      &-- &--    &--     &&-0.8&-0.7     &&--      &0.3&0.3&--      &--\medskip\\
TYC 2955$-$408$-$1  &0.03 (0.2) &--&0.43     &&0.13:      &--         &&0.6&0.52:&0.52:&0.52&0.52\\
      &--   &--  &--     &&0.7&0.6    &&--      &0.2&-0.2&--      &--\medskip\\
TYC 591$-$1090$-$1  &$-$0.07 (0.1)&-- &--     &&--      &--          &&0.9      &0.82:&0.82:&0.6 &-- \\
      &--  &--   &--     &&0.5&-0.4    &&--      &0.7&0.7&--      &--\medskip\\
\hline

\end{tabular}
}
\tablefoot{The first line for each star provides abundances from the present paper, and the second line those from N19,
which were converted from the original [X/Fe] data using 
N19 metallicity (as listed in Table~\ref{Tab:programme_stars}). [\ion{Sr}{I}/H] is the Sr abundance derived from \ion{Sr}{I} lines, and similarly for \ion{Sr}{II} and \ion{Ba}{II}. For the Sun, we adopted  $\log\epsilon_\odot$(Sr) = $2.87\pm0.07$ and $\log\epsilon_\odot$(Ba) = $2.18\pm0.09$, according to \citet{Asplund2009}. The value between parentheses in column [\ion{Sr}{I}/H] $\lambda$ 4607.327 is the NLTE correction according to \citet{Bergemann2012}, and the Sr abundance listed in that column is NLTE-corrected. For the criteria used to classify stars as 'no s-process', 'mild s-process enrichment' and 'strong s-process enrichment', see Sect.~\ref{Sect:classification}. 
Footnote (a): see discussion in Sect.~\ref{Sect:Individual}.}

\label{Tab:Sr}
\end{table*}

\subsection{Heavy s-process elements:  Ba, La, Ce}
\label{Sect:heavy-s}

We derived Ba abundances in most of the programme stars using the \ion{Ba}{II} lines at 5853.673 and 
6141.673 \AA. For a few objects, as these lines are strong and saturated, the Ba abundance is estimated from the spectral synthesis of the weak \ion{Ba}{II} line at 4524.924~\AA.
Ba lines are strongly affected by hyperfine (HF) splitting.
HF splitting data is not available for the 4524.924~\AA~  line, but it was
taken into account for the \ion{Ba}{II} 5853.673~\AA~ line.

\begin{figure}
\includegraphics[width=9cm]{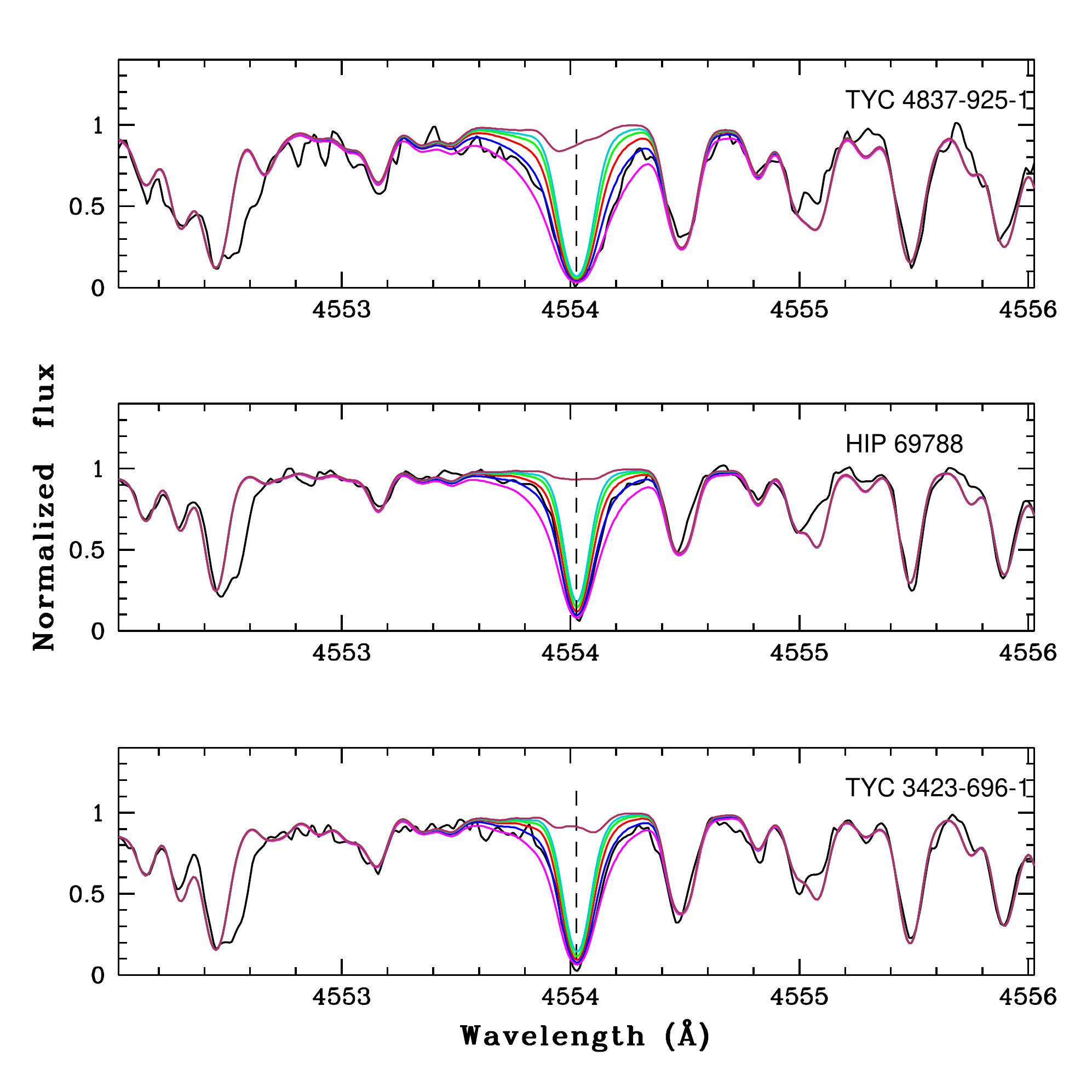}
\includegraphics[width=9cm]{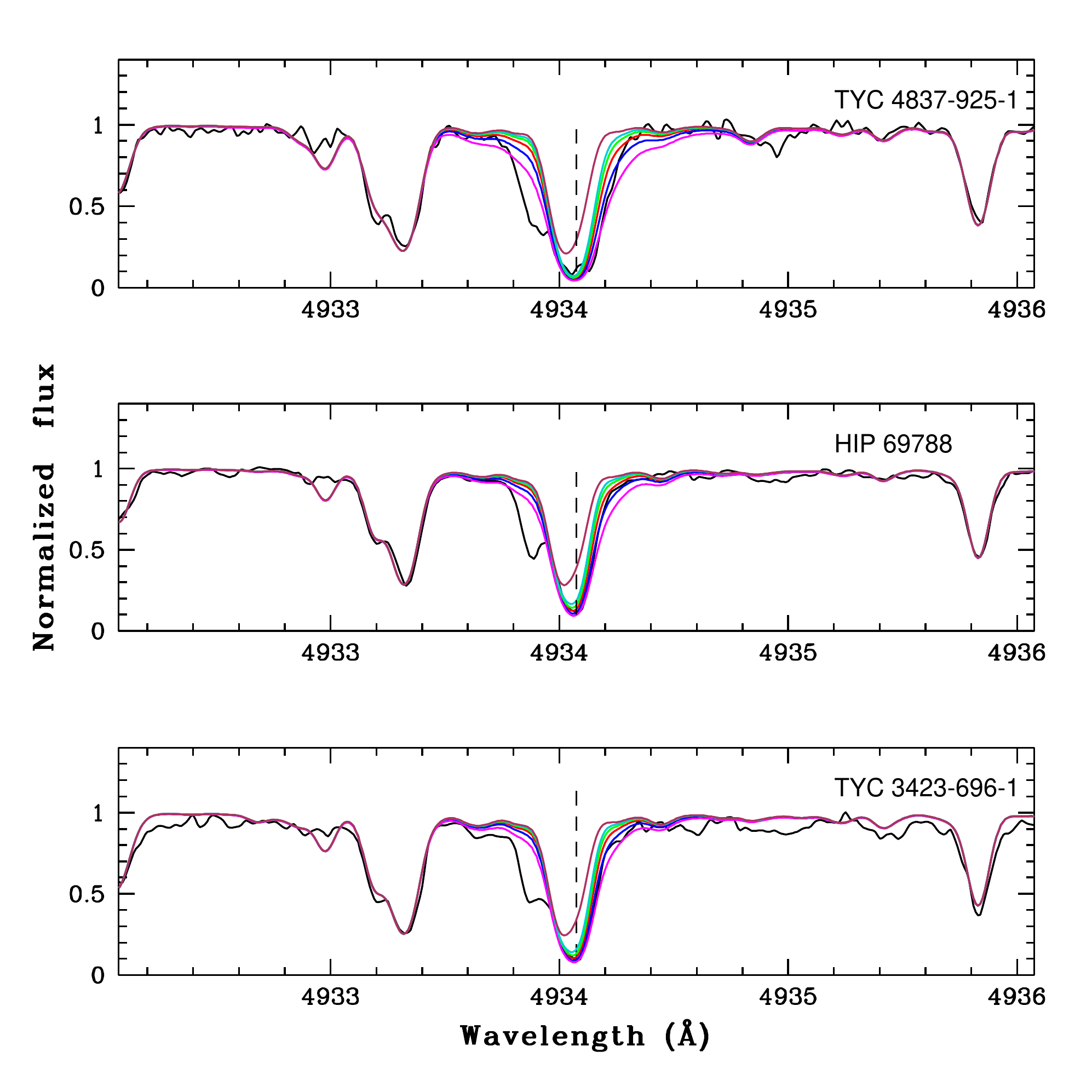}
\caption{The \ion{Ba}{II} 4554.026~\AA~ (top panel) and \ion{Ba}{II} 4934.076~\AA~ lines (bottom panel) are shown in three 
N19 Sr-only stars. The magenta, blue, red, green and cyan curves correspond to $\log \epsilon$(Ba) = 2.6, 2.3, 2.0, 1.7 and 1.4 dex respectively. The brown curve is for no Ba. }
\label{Fig:Ba-line}
\end{figure}

The \ion{Ba}{II} lines at 
4554.03 and 4934.08~\AA~ are saturated (Fig.~\ref{Fig:Ba-line}) and were therefore not considered to derive Ba abundances. Nevertheless, they were used by 
N19, and are listed in Table~\ref{Tab:Sr}. When comparing the derived Ba abundances, we conclude that the agreement between 
the N19 study and ours is much better for Ba than for Sr.

The La abundance is determined mainly using the lines for which HF splitting is available. 

As we mentioned earlier, all the lines used are listed in \citet{Karinkuzhi2018,Karinkuzhi2021} along with the hfs splitting and isotopic shifts as listed in these references.

\subsection{Abundance uncertainties}
Abundance  uncertainties are calculated for all elements using the methodology described in \citet{Karinkuzhi2018,Karinkuzhi2021}. Following Eq.~2 from \citet{Johnson2002},  the uncertainties on the elemental abundances $\log \epsilon$ write:
\begin{equation}
\begin{split}
\label{Eq:Johnson}
\sigma^{2}_{\rm tot}=\sigma^{2}_{\rm ran}
\;+\; \left(\frac{\partial \log\epsilon}{\partial T}\right)^{2}\sigma^{2}_{T}  \;+\; \left(\frac{\partial \log \epsilon}{\partial \log g}\right)^{2}\;\sigma^{2}_{\log g} \\
\;+\; \left(\frac{\partial \log \epsilon}{\partial \xi }\right)^{2}\;\sigma^{2}_{\xi}\;+\;
\left(\frac{\partial \log\epsilon}{\partial  \mathrm{[Fe/H]}}\right)^{2}\sigma^{2}_{\mathrm{[Fe/H]}} \;+\; \\ 2\bigg [\left(\frac{\partial \log\epsilon}{\partial T}\right) \left(\frac{\partial \log \epsilon}{\partial \log g}\right) \sigma_{T,\log g}
\;+\; \left(\frac{\partial \log\epsilon}{\partial \xi}\right) \left(\frac{\partial \log \epsilon}{\partial \log g}\right) \sigma_{\log g, \xi} \\
\;+\;\left(\frac{\partial \log\epsilon}{\partial \xi}\right) \left(\frac{\partial \log \epsilon}{\partial T}\right) \sigma_{ \xi, T}\Bigg],
\end{split}
\end{equation}
where $\sigma_{T}$, $\sigma_{\log g}$, and $\sigma_{\xi}$  are
the typical uncertainties on the atmospheric parameters and are derived by taking the average of the errors listed in Table~\ref{Tab:programme_stars} corresponding to each atmospheric parameter.  These values are estimated as $\sigma_{T}$ = 33~K, $\sigma_{\log g}$ = 0.26~dex, $\sigma_{\xi}$ = 0.06~km/s. The uncertainty on metallicity was estimated as  $\sigma_{\mathrm{[Fe/H]}}$ = 0.08~dex.
 The partial derivatives appearing in Eq.~\ref{Eq:Johnson} were evaluated in the specific cases of BD $-07^\circ$ 402, varying the
atmospheric parameters $T_{\rm eff}$, $\log g$, microturbulence $\xi$, and [Fe/H] by 100~K, 0.5, 0.5~km/s and 0.5 dex, respectively.
The resulting changes in the abundances are presented in Table~\ref{Tab:uncertainties}. The covariances $\sigma_{T,\log g}$, $\sigma_{\log g,\xi}$, and $\sigma_{\xi, T}$ are derived by the same method as given by \citet{Johnson2002}. In order to calculate  $\sigma_{T,\log g}$, we varied the temperature while fixing metallicity and microturbulence, and determined the $\log g$ value required for ensuring the ionization balance. Then using Eq.~3 of \citet{Johnson2002}, we derived the covariance $\sigma_{T,\log g}$ and found a value of 1.62. In a similar way, we found $\sigma_{\log g,\xi}$ = $-$0.02 and $\sigma_{\xi, T}$ = 0.75. 

The random error $\sigma_{\rm ran}$ is the line-to-line scatter.  For most of the elements, we could use more than four lines to derive the abundances. In that case, we have adopted $\sigma_\mathrm{ran} = \sigma_{l}/N^{1/2}$, where $\sigma_{l}$ is the standard deviation of the abundances derived from all the $N$ lines of the considered element. For the elements for which fewer number of lines are used to derive the abundances, we selected a $\sigma_{\rm ran}$ value as described in \citet{Karinkuzhi2021}. The final error on  [X/Fe] is derived from \\
\begin{equation}
\sigma^{2}_{\rm [X/Fe]} =\sigma^{2}_{\rm X} + \sigma^{2}_{\rm Fe} - 2\;\sigma_{\rm X,Fe},
\end{equation}
where $\sigma_{\rm X,Fe}$ is calculated using Eq.~6 from \citet{Johnson2002} with an additional term including $\left(\frac{\partial \log\epsilon}{\partial  \mathrm{[Fe/H]}}\right)$.

\begin{table}
\caption{Sensitivity of the abundances ($\Delta \log \epsilon_{X}$) with variations of the atmospheric parameters (considering the atmospheric parameters of BD $-07^\circ$ 402 ). }
\label{Tab:uncertainties}
\begin{tabular}{crrrr}
\hline
\\
       & \multicolumn{4}{c}{$\Delta \log \epsilon_{X}$} \\
        \cline{2-5}
        \\
 Element&   $\Delta T_{\rm eff}$ & $\Delta \log g$  &$\Delta$ [Fe/H]& $\Delta \xi_t$   \\
&   ($+$100 K) & ($+$0.5) & ($+$0.5  & ($+$0.5 \\
   &        &         & dex)& km~s$^{-1}$) \\
\hline\\
Li & 0.15 & 0.00  &   0.00 &   0.00\\
C  & 0.00  & 0.15 &  0.10  &  0.00 \\
N  & 0.10  & 0.30  &  0.30  & 0.10	\\
O  & 0.00  & 0.20  &  0.15 &  0.00	 \\
Na & 0.11 & 0.08 & $-$0.15 & $-$0.05  \\
Fe & 0.13 & 0.25 &  0.20 & $-$0.15 \\
Rb & 0.00  & 0.05 & $-$0.05 &  0.00\\
Sr & 0.30  & 0.15 &  0.15 & 0.00   \\
Y  & 0.05 & 0.13 &  0.03 &  0.25   \\
Zr & 0.03 & 0.07 & $-$0.08 & $-$0.05  \\
Ba & 0.05 & 0.17 & $-$0.15 & $-$0.40   \\
La & 0.00  & 0.16 &  0.05 &  0.04 \\
Ce & 0.04 & 0.20 & 0.08  & $-$0.05 \\
\hline
\end{tabular}
\end{table}

\section{Classification based on abundance ratios}
\label{Sect:classification}

Based on the abundances listed in Table~\ref{Tab:abundances}, we classify our programme stars according to the following criteria:
\begin{itemize}
\item[-] "no": all heavy elements listed in Table~\ref{Tab:abundances} have [X/Fe]~$ < 0.2$;
\item[-] "mild": at least three heavy elements are in the range $0.2 < $~[X/Fe]~$ < 0.8$;
\item[-] "strong": at least three  heavy elements have [X/Fe]~$ \ge 0.8$.
\end{itemize}

\begin{figure}
\includegraphics[width=9cm]{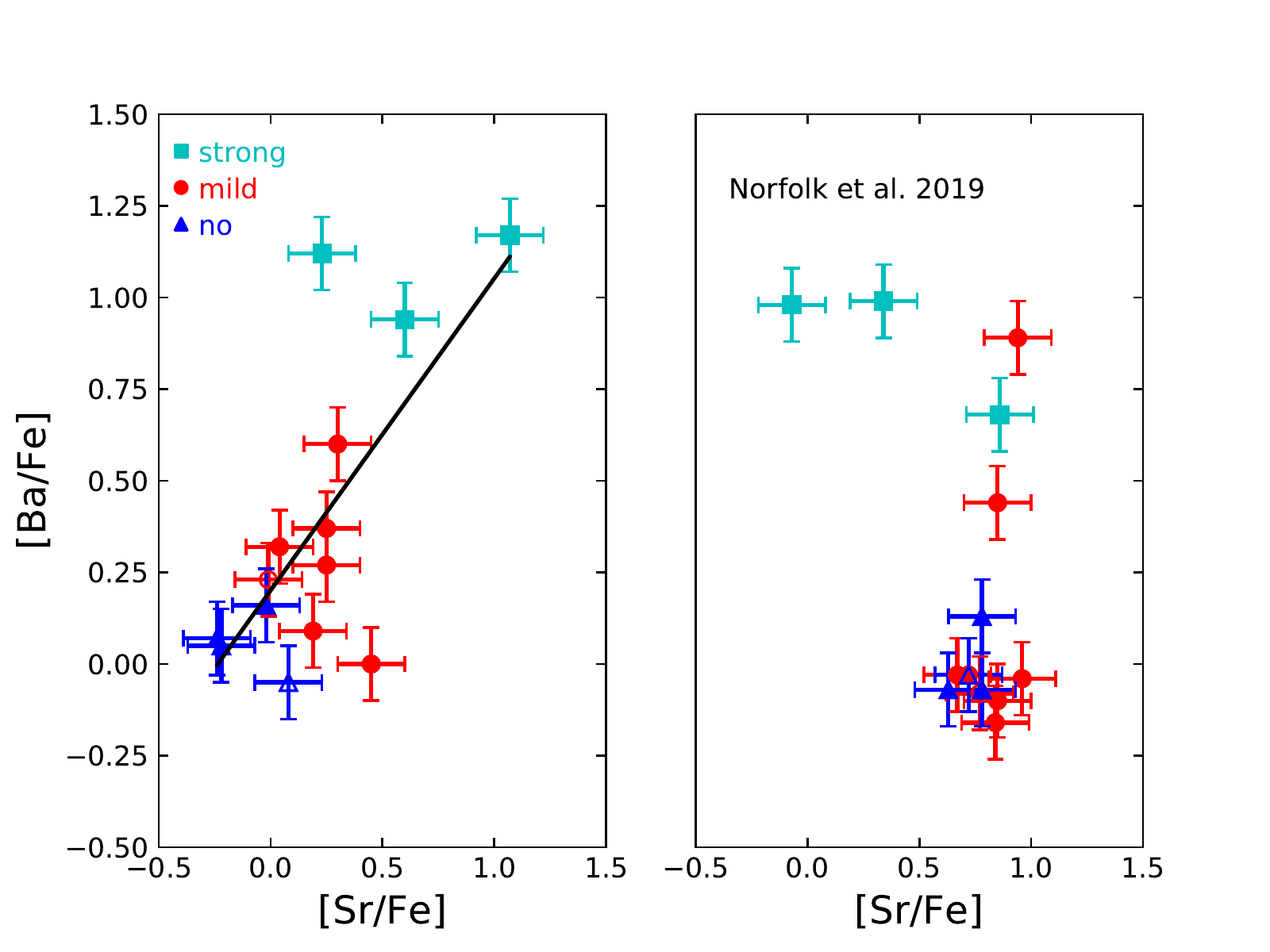}
\caption{
The ([Ba/Fe], [Sr/Fe]) plane from this analysis (left panel) and from N19 (right panel). The left panel reveals that there exists a correlation between the [Sr/Fe] and [Ba/Fe]  abundance ratios,} represented by the solid line corresponding to a least-square fit to the data. The "strong", "mild" and "no" stars
(see Sect.~\ref{Sect:classification}) are color-coded as indicated in the label.
Open symbols refer to stars with a  Li abundance determination. The right panel shows [Ba/Fe] and [Sr/Fe] from N19.
\label{Fig:BaFe}
\end{figure}

As shown in Fig.~\ref{Fig:BaFe}, Sr and Ba abundances are correlated (left panel), with just one star (TYC 3423$-$6966$-$1)  falling the farthest away from the regression line, with a marginal Sr excess ([Sr/Fe]~$ 
\sim 0.5$) and no Ba excess ([Ba/Fe]~$\sim 0$). Our analysis thus finds no "Sr-only" stars (which would correspond to stars with only Sr overabundant -- or more generally with only the first-s-process-peak elements  overabundant, which are not present in Table~\ref{Tab:abundances}), as opposed to the 13 stars flagged as such by 
N19 (see bottom right panel of Fig.~\ref{Fig:BaFe} and Table~\ref{Tab:programme_stars}). The 13 
"Sr-only" stars of N19 split in 4 "no-s" stars, 8 mild barium stars, and 1 strong barium star.

\begin{figure}
\includegraphics[width=9cm]{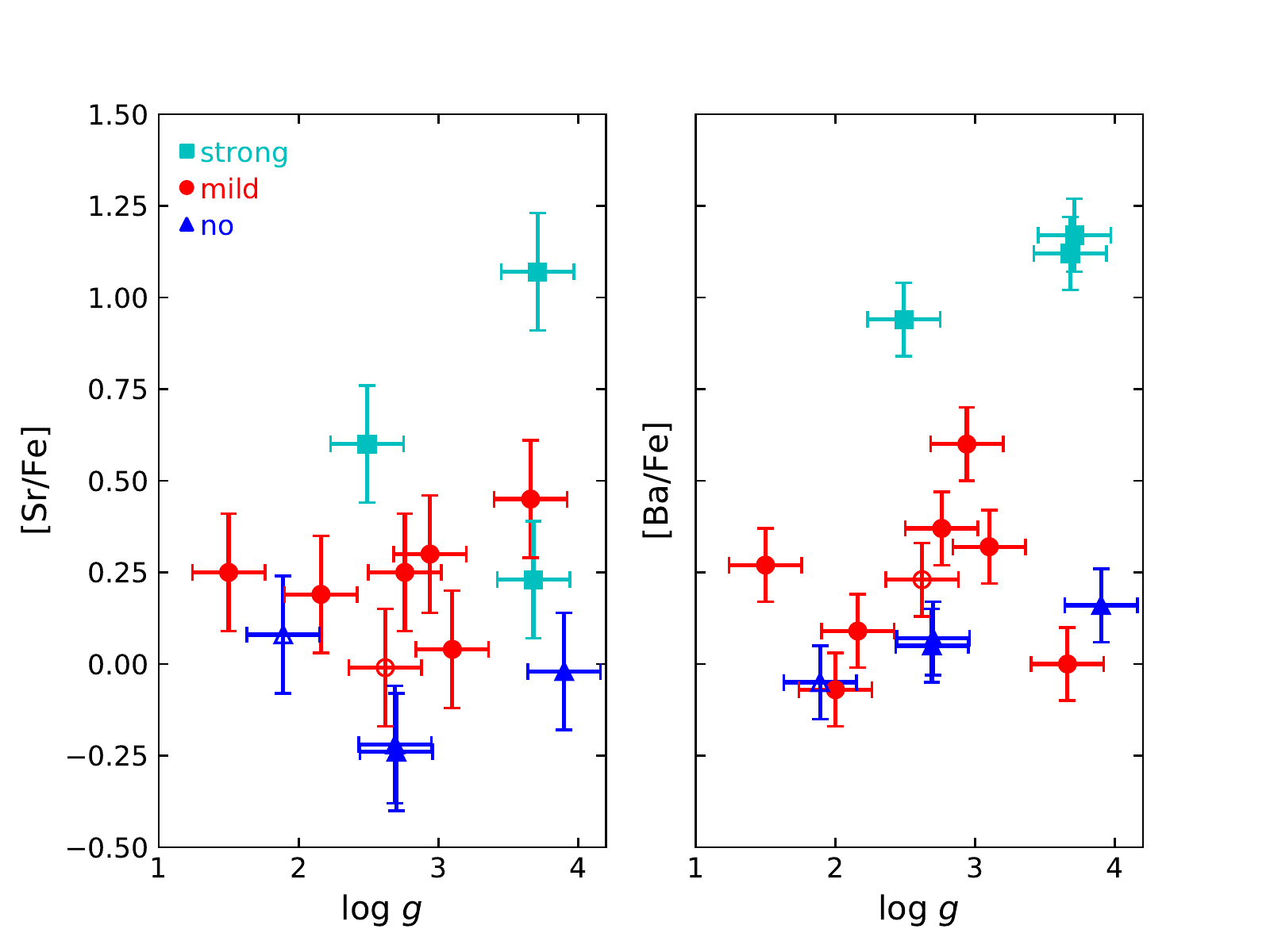}
\caption{The [Sr/Fe] and [Ba/Fe] abundance ratios vs $\log g$. 
Symbols are as in Fig.~\ref{Fig:BaFe}.
}
\label{Fig:SrBalogg}
\end{figure}

Figure~\ref{Fig:SrBalogg}
reveals as well that "no-s" stars  {\it cannot} be attributed to an unrecognized positive luminosity effect on the \ion{Sr}{II} lines (i.e., a low gravity $\log g$ causing a strengthening of lines from ionized species), since these "no-s" stars are not restricted to low $\log g$ values.

\begin{figure}
\includegraphics[width=9cm]{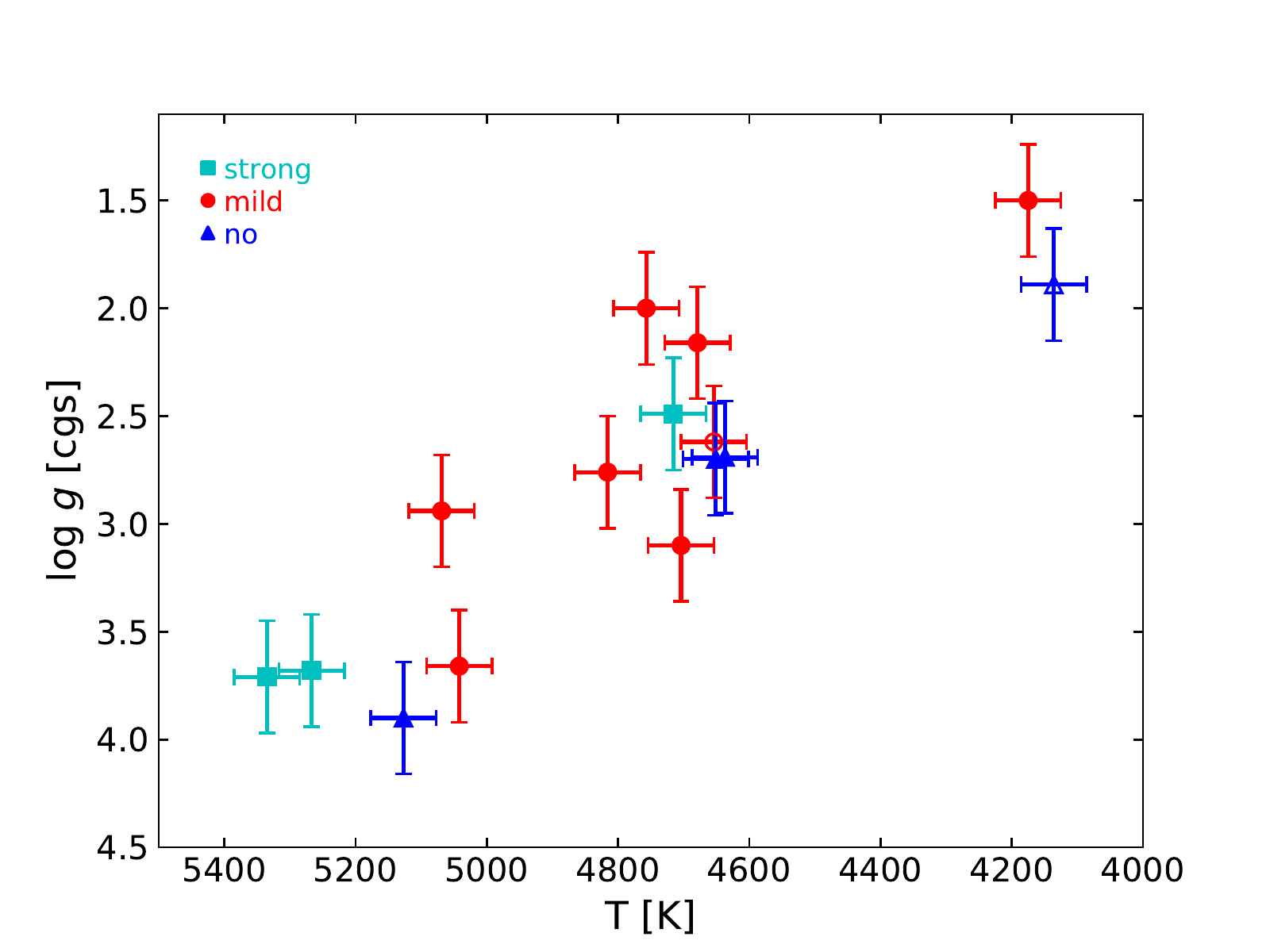}
\caption{Kiel diagram ($\log g$ vs $T_{\rm eff}$) for the programme stars.}
\label{Fig:Tefflogg}
\end{figure}

It is worth noting that the two most enriched barium stars are actually barium dwarfs
(Figs.~\ref{Fig:SrBalogg} and \ref{Fig:Tefflogg}).
Curiously enough, N19
%Norfolk at al. 
mention that, because their machine-learning algorithm used a training sample composed of giant stars only, they are therefore "unable to identify s-process enhanced dwarf stars". Here we show the contrary, because the 4554~\AA~  and 4934~\AA~ 
\ion{Ba}{II} lines are sensitive to a Ba enhancement both in giants and dwarfs (in Fig.~\ref{Fig:Ba-line}, HIP~69788 and TYC~3423-696-1 are dwarfs whereas TYC~ 4837-925-1 is a giant). Therefore, the residuals between the observed flux at those wavelengths and the Cannon data-driven model will be able to identify barium stars, irrespective of them being dwarfs or giants.

\section{Discussion of individual stars}
\label{Sect:Individual}

{\bf BD $\mathbf{-07^\circ402}$}: This object is the only mild Ba star  with a measurable Li abundance of  $\log \epsilon_{\rm Li} = 1.3$~dex, just large enough to qualify it as a Li-rich K giant \citep{Jorissen2020}. \\
{\bf BD $\mathbf{+44^\circ575}$}: 
This mild Ba star presents strong enrichments in Na and Mg.  \\
{\bf HD 7863}: 
This "no-s" star is one of four objects in our sample that exhibits a larger than average N abundance ([N/Fe]~$ \ge 0.7$; Fig.~\ref{Fig:CNBa} and Sect.~\ref{Sect:CNO}).\\
{\bf HIP 69788}: 
 This star has atmospheric parameters which differ the most between our study and that of
 N19 (see Table~\ref{Tab:programme_stars} and Sect.~\ref{Sect:TheCannon}). \\
{\bf TYC 22$-$155$-$1}: 
This star exhibits a strong enrichment in Mg. \\
{\bf TYC 2913$-$1375$-$1}: For this star, an accurate Sr abundance could not be derived, since the \ion{Sr}{I} lines are weak and the spectrum is too noisy in the violet to access the \ion{Sr}{II} lines.
Nevertheless, Zr, Ba, and La lines reveal that this star has mild s-process enhancements.\\
{\bf TYC 3144$-$1906$-$1}: This is the second object with a measured Li abundance, but not large enough though ($\log \epsilon_{\rm Li} = 0.6$~dex) to be considered as a Li-rich K giant. While it is not enriched in any  s-process elements, this star shows a high enhancement in N ([N/Fe]~=~1.40~dex; Fig.~\ref{Fig:CNBa}).  \\
{\bf TYC 4684$-$2242$-$1}: Another "no-s" star with a larger than average N abundance (Fig.~\ref{Fig:CNBa}). It is worth noting that 3 out the 4 N-rich stars ([N/Fe]~$\ge 0.7$) do not show any s-process enhancement. \\

\section{Origin of the peculiarities of mild and strong barium stars}
\label{Sect:originofpattern}

\subsection{Binary frequency} 

According to the canonical scenario  \citep{McClure1983,Jorissen2019,Escorza2019}, barium stars form in a binary system. 
Table~\ref{Tab:kinematics} therefore collects the kinematical properties of our programme stars. Contrary to what is announced in their paper, 
N19
do not provide the LAMOST radial velocities (RVs) in their supplementary material. Therefore, we resorted to Gaia Data Release 2 \citep[GDR2;][]{Katz2019} to get one RV value to compare to the HERMES value. GDR2 velocities refer to epoch 2015.5 (JD 2457205), whereas HERMES RVs were taken roughly 1700~d later, offering a large enough time span to efficiently detect even long-period binaries. 

The uncertainty $\epsilon$ (listed in Table~\ref{Tab:kinematics}) on the GDR2 RVs has been computed from $G_{\rm RVS}$ (the Gaia magnitude in the RVS band) along the same method as discussed by \citet[][their Eqs.~3 and 4]{Jorissen2020}, except that Eq.~(4) has been replaced by one depending on the Tycho $B_T - V_T$ color index, as listed in Table~7 of \citet{Jordi2018}. As can be seen in Table~\ref{Tab:kinematics}, uncertainties on the GDR2 RVs are on the order of 0.3~km~s$^{-1}$ for the brightest targets ($G_{\rm RVS} \sim 8$) and go up to 1~km~s$^{-1}$ for the faintest objects ($G_{\rm RVS} \sim 10.8$). We note that most of the stars have the computed $\epsilon$ uncertainty is very similar to the RV uncertainty listed by GDR2 (except for the binary HIP~69788).
Based on the above, we have flagged as binaries all stars with $\Delta RV \ge 3\;\epsilon$. 

As expected, 2 out of 3 strong barium stars show a clear binary signature, and the third one (TYC~591$-$1090$-$1) is the faintest in the sample, and despite its large $\Delta RV$ value of 1.55~km~s$^{-1}$, it does not fulfill the $3\epsilon$ condition. The results for the other classes are intriguing, 
since only one of the 7 mild barium stars diagnosed exhibit statistically-significant RV variations, and on the opposite, 2 out of the 4 'no-s' stars show a binary signature.

We see no obvious explanation for the high prevalence of binaries among the latter category, other than small-number fluctuations. Concerning the mild Ba stars, it is still possible that their RV variations are more difficult to detect since they contain a larger proportion of binaries with very long periods ($> 5000$~d) than strong barium stars do  \citep[see Fig.~7 of][]{Jorissen2019}, thus explaining the lower apparent frequency of binary signatures among mild barium stars.
Alternatively, some mild barium stars with no binary signature might represent the upper tail of the [Ba/Fe] range 
in the Galactic ([Ba/Fe], [Fe/H])
trend \citep{Edvardsson1993}. Such mild barium stars are especially present
at metallicities in the range $[-0.4\; +0.1]$~dex
(Fig.~15 panel $k$ of \citealt{Edvardsson1993}, Fig.~5 of \citealt{Tautvaisiene-2021}).

\begin{figure}
\includegraphics[width=9cm]{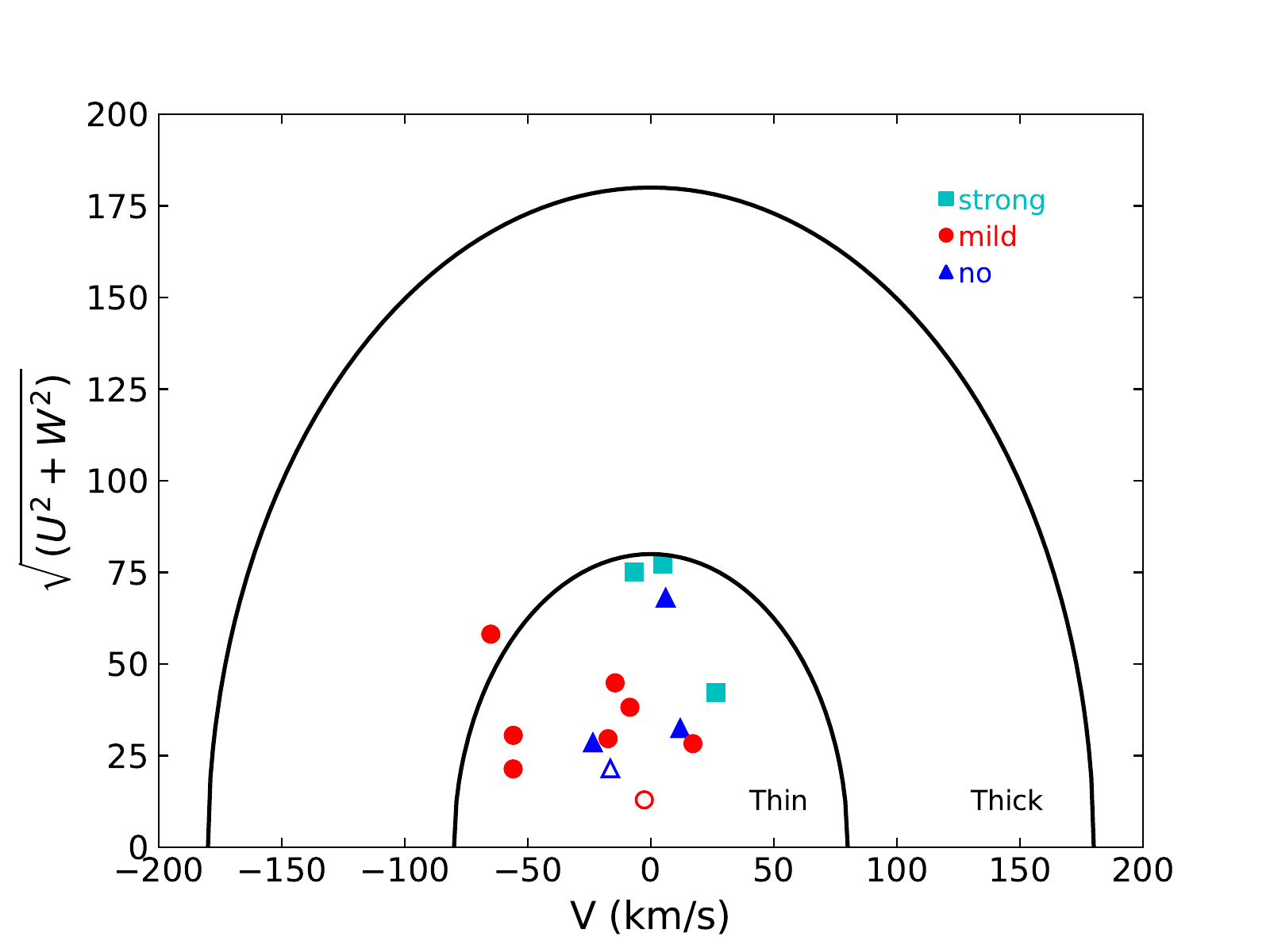}
\caption{Toomre diagram of the programme stars, with the black line delineating the location of thin-disk stars (as in Fig.~4 of N19). }
\label{Fig:Toomre}
\end{figure}

All non-binary mild barium stars but TYC~2913-1375-1 have [Fe/H] in the above range, making it likely 
that they owe their mild-barium nature to the fluctuations in the Galactic chemical evolution.
 Figure~\ref{Fig:Toomre}, which displays the Toomre diagram of the programme stars, reveals no difference in their kinematic properties, they all belong to the Galactic thin disc.
  
\setlength{\tabcolsep}{2pt}
\begin{table*}\small
\caption{Kinematic properties of the programme stars. 
}
\label{Tab:kinematics}
\begin{tabular}{llrrrrrrrrrc}
\hline
\\
Name &  \multicolumn{1}{c}{$JD$}&\multicolumn{1}{c}{RV}  & \multicolumn{1}{c}{$\Delta RV$} & \multicolumn{1}{c}{$\epsilon$}& \multicolumn{1}{c}{$G_{\rm RVS}$}& \multicolumn{1}{c}{source}& \multicolumn{1}{c}{$U$} & \multicolumn{1}{c}{$V$}& \multicolumn{1}{c}{$W$} & \multicolumn{1}{c}{Gaia DR2} & Bin./Rem.\\
      & (-2400000) & \multicolumn{1}{c}{(km s$^{-1})$} & \multicolumn{1}{c}{(km s$^{-1})$} & \multicolumn{1}{c}{(km s$^{-1})$}&&&\multicolumn{1}{c}{(km s$^{-1})$} & \multicolumn{1}{c}{(km s$^{-1})$} & \multicolumn{1}{c}{(km s$^{-1})$}\\
\hline\\
\multicolumn{11}{c}{\bf no s-process enrichment}\medskip\\
HD 7863& 57205 & $-27.82\pm0.34$ & 0.45&0.29 & 8.09 & GDR2 & 24.95 &	-23.58 &	-13.81 & 397523176280439808 & n\\
       & 58882.41& $-28.27\pm0.07$ &&&& HER\medskip\\
       HIP 69788& 57205& $6.56\pm1.94$& 55.93 &0.42& 9.39& GDR2 &31.08& 11.94& -9.27 &3667671452515762944& y \tablefootmark{a}\\
 & 58877.70 & $-49.37\pm0.07$ &&&& HER\medskip\\
TYC 3144$-$1906$-$1& 57205& $-21.56\pm0.33$ & 1.06 &0.28&7.84&GDR2 &-20.73 & -16.47& -5.61 & 2077143186195922176 & y
\\
 & 59090.45 & $-22.62\pm0.07$ &&&&HER\medskip\\
TYC 4684$-$2242$-$1 & 57205& $36.53\pm0.44$ & 0.92 &0.41&9.35& GDR2  &-66.36 & 6.01 &  -15.01 &  2478965826587133568 & n\\
 & 58882.33 & $35.61\pm0.07$ &&&& HER\medskip\\
\multicolumn{11}{c}{\bf mild s-process enrichment}\medskip\\
BD $-07^\circ$ 402 & 57205& $-12.96\pm0.17$ & 0.48 &0.29&8.04& GDR2 &4.72& -2.66& 12.05 & 2486894817251498240&n \\
 & 58882.37 & $-13.44\pm0.07$ &&&& HER\medskip\\
BD $+44^\circ$ 575 & 57205& $-12.61\pm0.26$ & 0.50 &0.28& 7.78&GDR2 &28.21& 17.09& 1.94 &  340768207120632832&n\\
  & 58878.41 & $-13.11\pm0.07$ &&&& HER\medskip\\
TYC 22$-$155$-$1& 57205&$-37.70\pm 0.58$ & 0.35 &0.33&8.71& GDR2 & -48.38& -65.05& 32.26&  2539172197106047872&n \\
 & 58882.36 & $-38.05\pm0.07$ &&&& HER\medskip\\
TYC 2913$-$1375$-$1&57205 & $-2.89\pm0.42$ & 0.15 &0.48& 9.70 & GDR2 &2.22 &-14.52 & 44.81 & 193703372946249216&n \\
 & 59090.71 & $-3.04\pm0.07$ &&&& HER\medskip\\
TYC 3305$-$571$-$1& 57205& $-57.93\pm 0.33$ & 0.52 &0.37&9.08& GDR2 &30.31& -55.92& 3.80  & 437946515118550016&n\\
 & 58878.48 & $-58.45\pm0.07$ &&&& HER & & & & \medskip\\
 TYC 4837$-$925$-$1& 57205& $-22.91\pm0.23$ & 0.52 &0.37& 9.11&GDR2 & 35.55& -8.52& -13.98& 3081162263449705984& n\\
  & 58877.55& $-23.43\pm0.07$&&&& HER\medskip\\
TYC 3423$-$696$-$1& 57205& $5.37\pm0.21$ & $-3.16$ &0.40& 9.26&GDR2 &-21.20 & -55.96& -2.78& 1016739606459940608 &y \\
 & 58877.61 & $8.53\pm0.07$&&&&HER\medskip\\
 TYC 752$-$1944$-$1& 58877.49 & $19.04\pm0.07$ &- &&& HER &4.66& -17.38& -29.26   &3157928756551254400 & ?
\medskip\\
\multicolumn{11}{c}{\bf strong s-process enrichment}\medskip\\
TYC 2250$-$1047$-$1& 57205& $19.39\pm0.75$ & 3.44 &0.75& 10.41&GDR2 & -76.78 &4.84 & -9.30& 2839977000550809856 &y\\
 & 59090.61 & $15.95\pm0.07$ &&&& HER\medskip\\
TYC 2955$-$408$-$1& 57205 & $41.15\pm0.56$ & 2.07 &0.42& 9.41& GDR2 &-61.94& -6.77& -42.54 &  953203601197511808 &y\\
 & 58879.54& $39.08\pm0.07$ &&&& HER\medskip\\
TYC 591$-$1090$-$1&57205 & $8.80 \pm0.88$ & 1.55 &1.00& 10.83&GDR2 &41.23& 26.56& 9.10 &  2757528128276067840 &n \\
 & 59090.64 & $7.25\pm0.07$ &&&& HER\\
\hline
\end{tabular}
\tablefoot{In column 'source', GDR2 stands for Gaia Data Release 2, and 'HER' for HERMES. $\Delta RV$ is the difference between the HERMES and GDR2 radial velocities. $\epsilon$ is the uncertainty on the GDR2 RV, computed as explained in the text. $G_{\rm RVS}$ is the magnitude in the RVS band \citep{Jordi2018}. $U, V, W$ are the Cartesian components of the velocity in Galactic coordinates, taken from
N19.
The Gaia DR2 identifier has been listed to ease the cross match with N19
data table. For the criteria used to classify stars as 'no s-process', 'mild s-process enrichment' and 'strong s-process enrichment', see Sect.~\ref{Sect:classification}.\\
\tablefoottext{a}{Also proper motion anomaly \citep{Kervella2019}.}
}
\end{table*}

\subsection{Location in the HRD} 

\begin{figure}
\includegraphics[width=8cm]{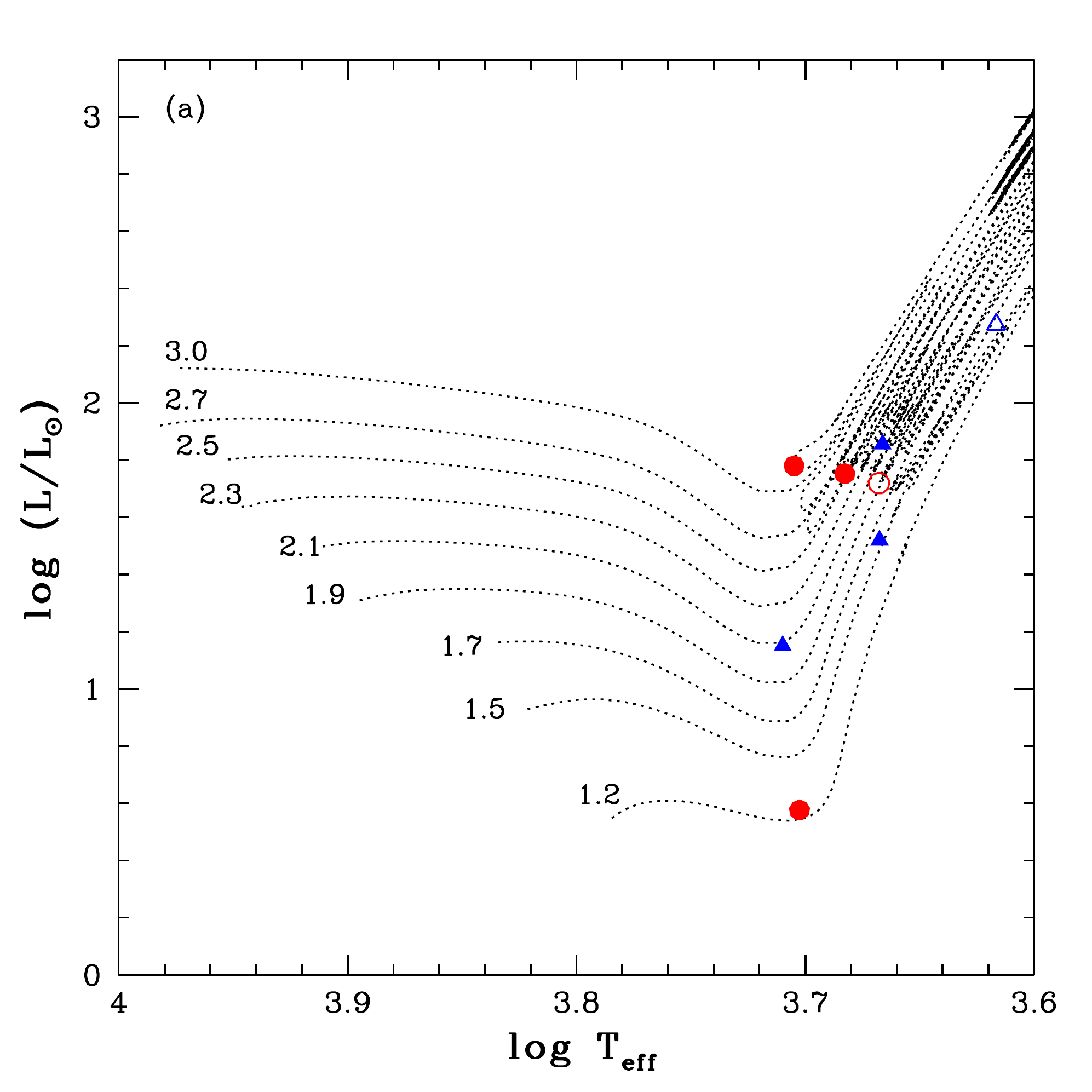}
\includegraphics[width=8cm]{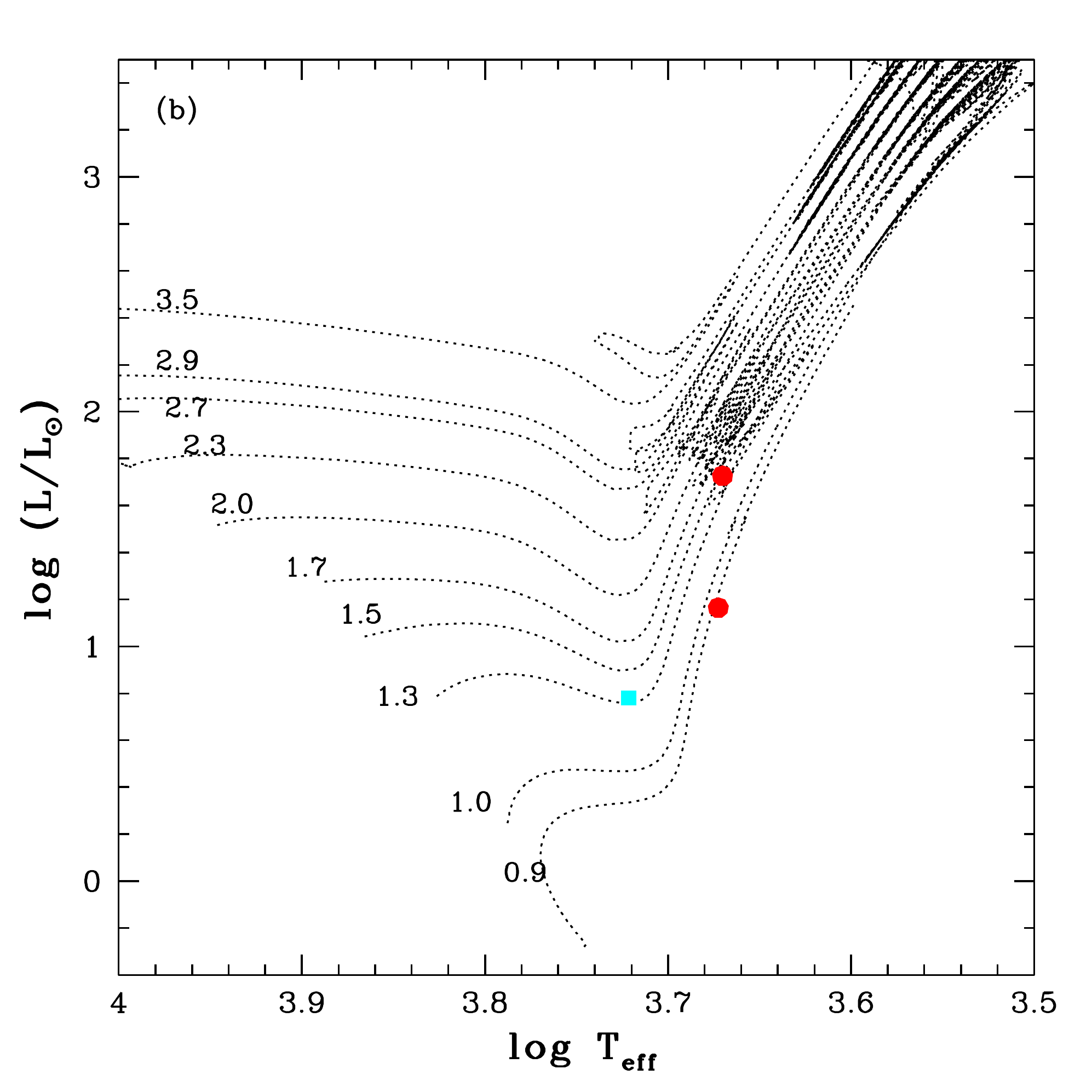}
\includegraphics[width=8cm]{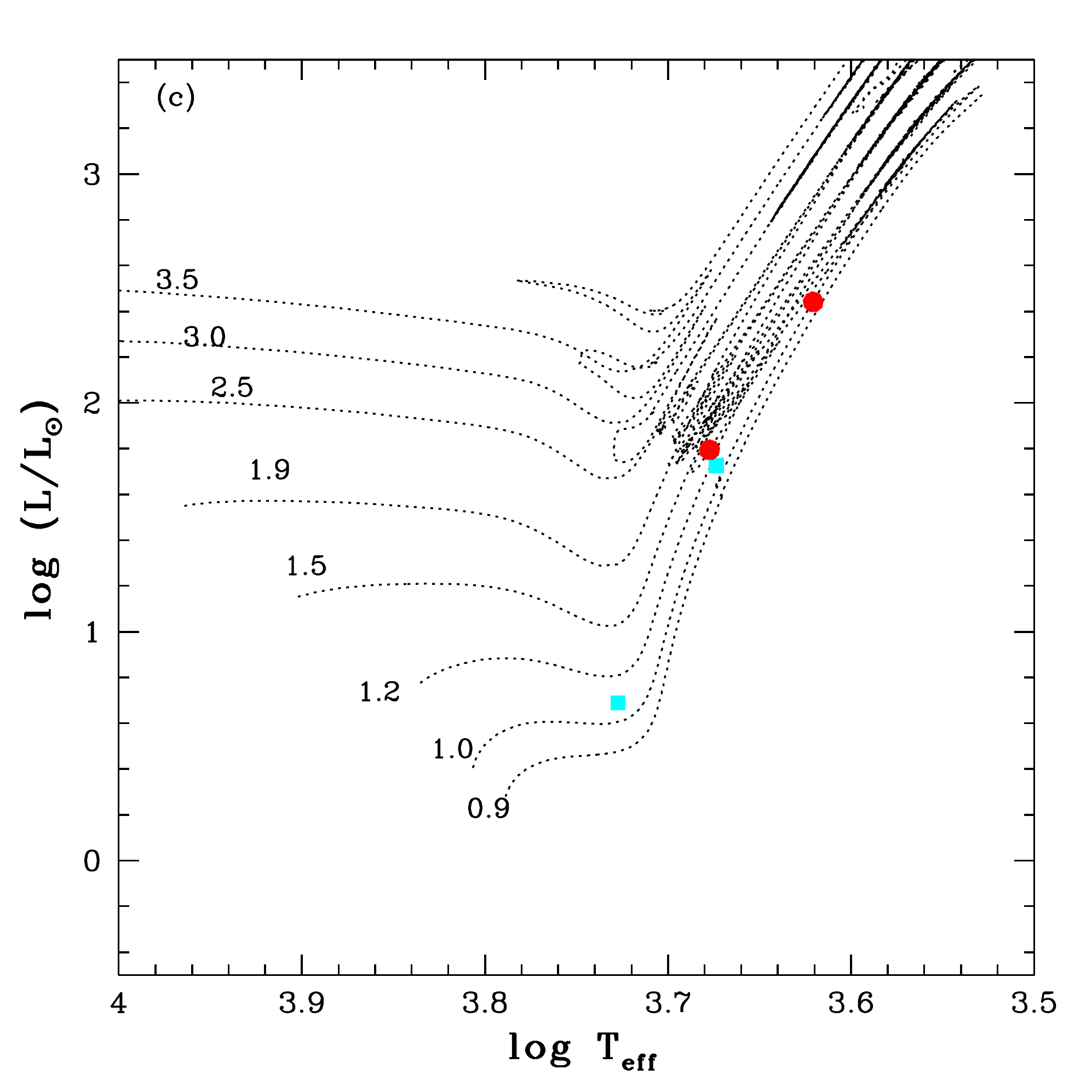}
\caption{Location of barium stars in the HRD compared with STAREVOL tracks: (a) [Fe/H] = 0  (b)  [Fe/H] $= -0.25$ (c)  [Fe/H] $= -0.5$. Symbols are as in Fig.~\ref{Fig:Sr-N}}
\label{Fig:HRD}
\end{figure}

Figure~\ref{Fig:HRD}  presents the location of our programme stars in the Hertzsprung-Russell diagram (HRD). In this plot, we used the spectroscopic $T_{\rm eff}$ values listed in Table~\ref{Tab:programme_stars} and determined the luminosity of each target by combining the flux obtained from integrating their spectral energy distributions (SED) with the distances computed by \citet{Bailerjones2021} from the Gaia Early Data Release 3 parallaxes \citep{eGDR32020}. To build and fit the SEDs, we applied the methodology described by \citet{Escorza2017} and successfully used in combination with spectroscopic parameters. The tool performs a $\chi^{2}$-grid-search to find the best-fitting MARCS model atmosphere \citep{Gustafsson2008} to the available broadband photometry for each target collected from the SIMBAD database \citep{Wenger2000}, treating the total line-of-sight reddening $E_{B - V}$ as a free parameter for which we optimise. We used the parameter ranges obtained from the spectroscopic analysis to limit $T_{\rm eff}$ and $\log g$, and we fixed the metallicity to the closest available in the MARCS grid ($0.0, -0.25$ or $-0.5$ for our targets), leaving $E_{B - V}$ as the only fully unconstrained parameter. Then each best-fitting SED model is corrected for interstellar extinction assuming that the line-of-sight extinction $A_{V}$ follows the Galactic extinction law given by $A_{V} = R_{V} \times E_{B - V}$ with $R_{V} = 3.1$ \citep{Fitzpatrick1999}. The SED is then integrated to get the total flux.

The star locations in the HRD are compared with evolutionary tracks from STAREVOL \citep{Siess2008} for three metallicities, [Fe/H] = $-0.5, -0.25$, and 0. A
correlation  may seem to exist between mass and metallicity: at the lowest metallicity [Fe/H]$ = -0.5$ (panel a of Fig.~\ref{Fig:HRD}), barium stars are found in the full mass range 1.2 - 3 M$_\odot$, whereas at solar metallicity (panel c of Fig.~\ref{Fig:HRD}), they are restricted to the much narrower range 0.9 -- 1.5~M$_\odot$.  This correlation is not confirmed, however, by the larger sample studied by \citet[][their Fig.~ 17]{Jorissen2019}. Thus the segregation observed in Fig.~\ref{Fig:HRD} is likely the result of small-sample statistics. 

To summarize, the confirmed barium stars from N19 are found in the mass range 0.9 -- 3~M$_\odot$ and are located all the way from the end of the main-sequence till the red clump through the red-giant phase.

\subsection{Abundance trends}

\begin{figure}
\includegraphics[width=9cm]{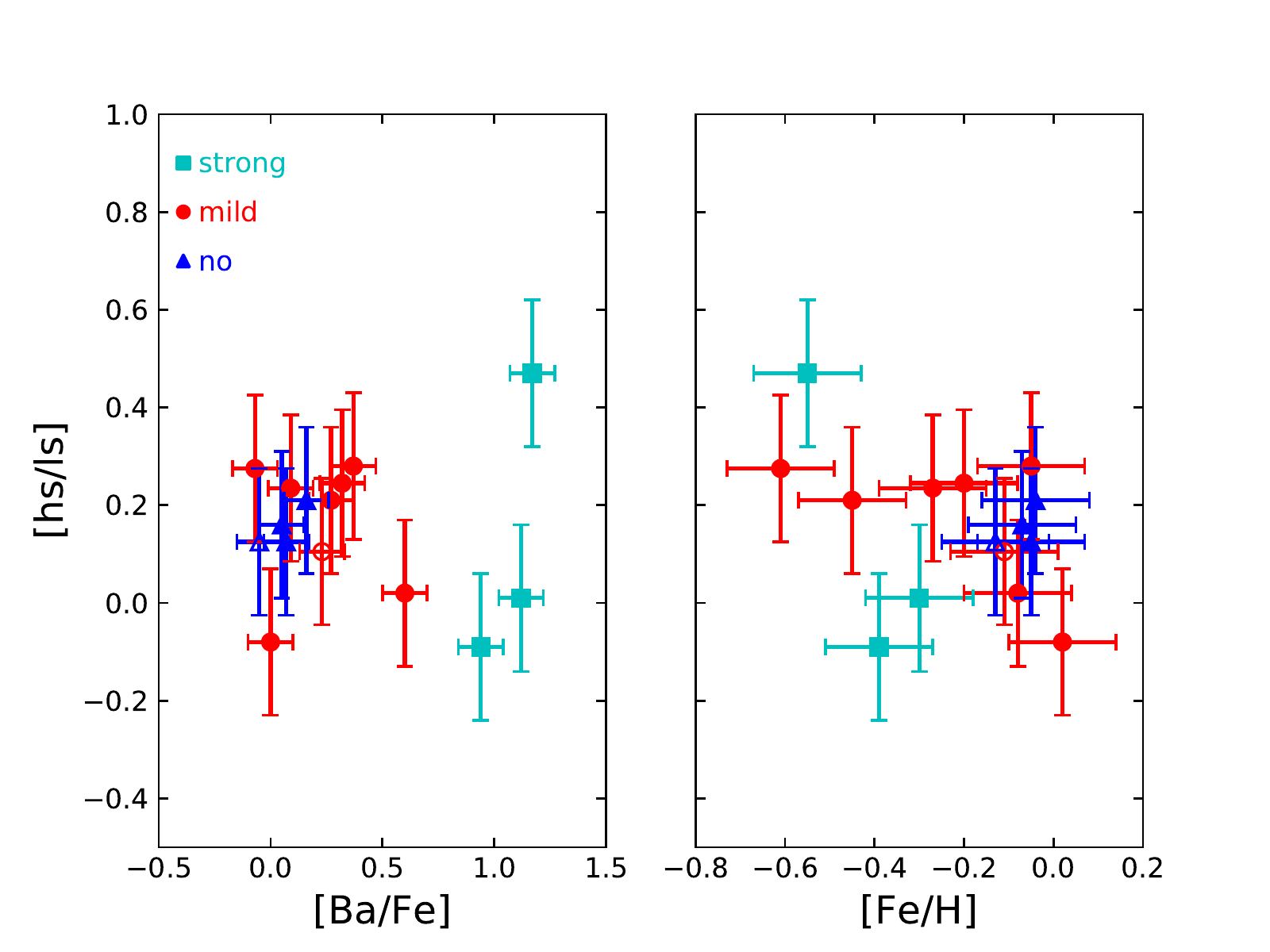}
\caption{The [hs/ls] ratio as a function of various abundance ratios.}
\label{Fig:hslsBa}
\end{figure}

In most of the barium stars studied here, the second s-process peak reaches slightly larger  overabundance levels than the first peak, resulting in  [hs/ls] ratios ranging from 0 to 0.5 dex (left panel of Fig.~\ref{Fig:hslsBa}), with only two exceptions (TYC 3423-696-1 and TYC 2955-408-1) where [hs/ls]~$ = -0.1$.    As usual in barium stars, the [hs/ls] ratio does not show a strong correlation with metallicity (right panel of Fig.~ \ref{Fig:hslsBa}).

\begin{figure}
\includegraphics[width=9cm]{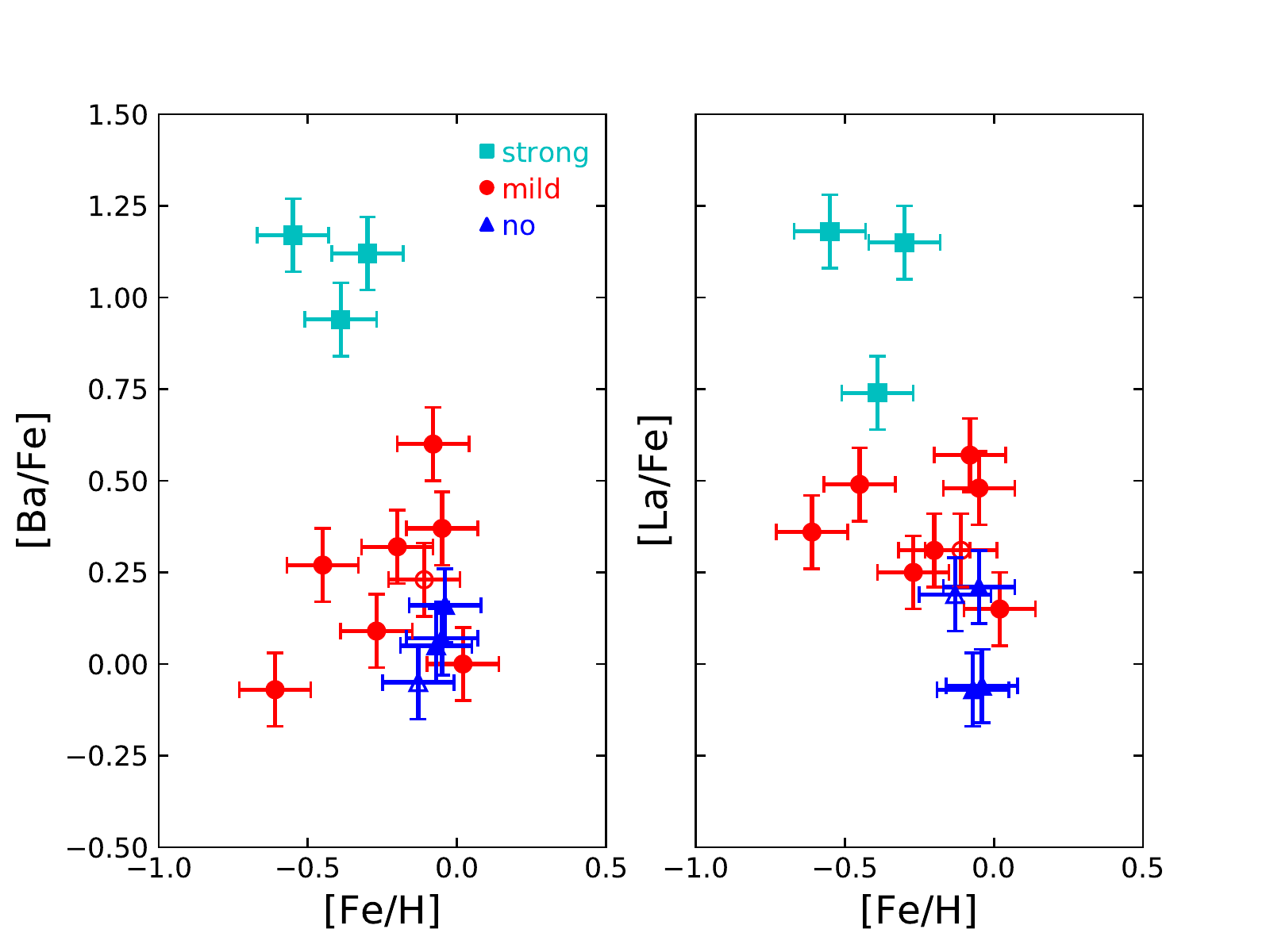}
\caption{The [Ba/Fe] and [La/Fe] ratios as a function of metallicity [Fe/H].}
\label{Fig:BaLaFe}
\end{figure}

The distribution of [La/Fe] vs [Fe/H]  (right panel of Fig.~\ref{Fig:BaLaFe}) indicates that there might be a weak correlation between metallicity and  the level of  s-process enrichment, since strong barium stars with [La/Fe]~$\ge 0.75$ have [Fe/H]~$\le -0.30$ whereas mild barium stars cluster instead around a slightly subsolar metallicity ([Fe/H]~$\sim -0.1$). 

\begin{figure}
\includegraphics[width=9cm]{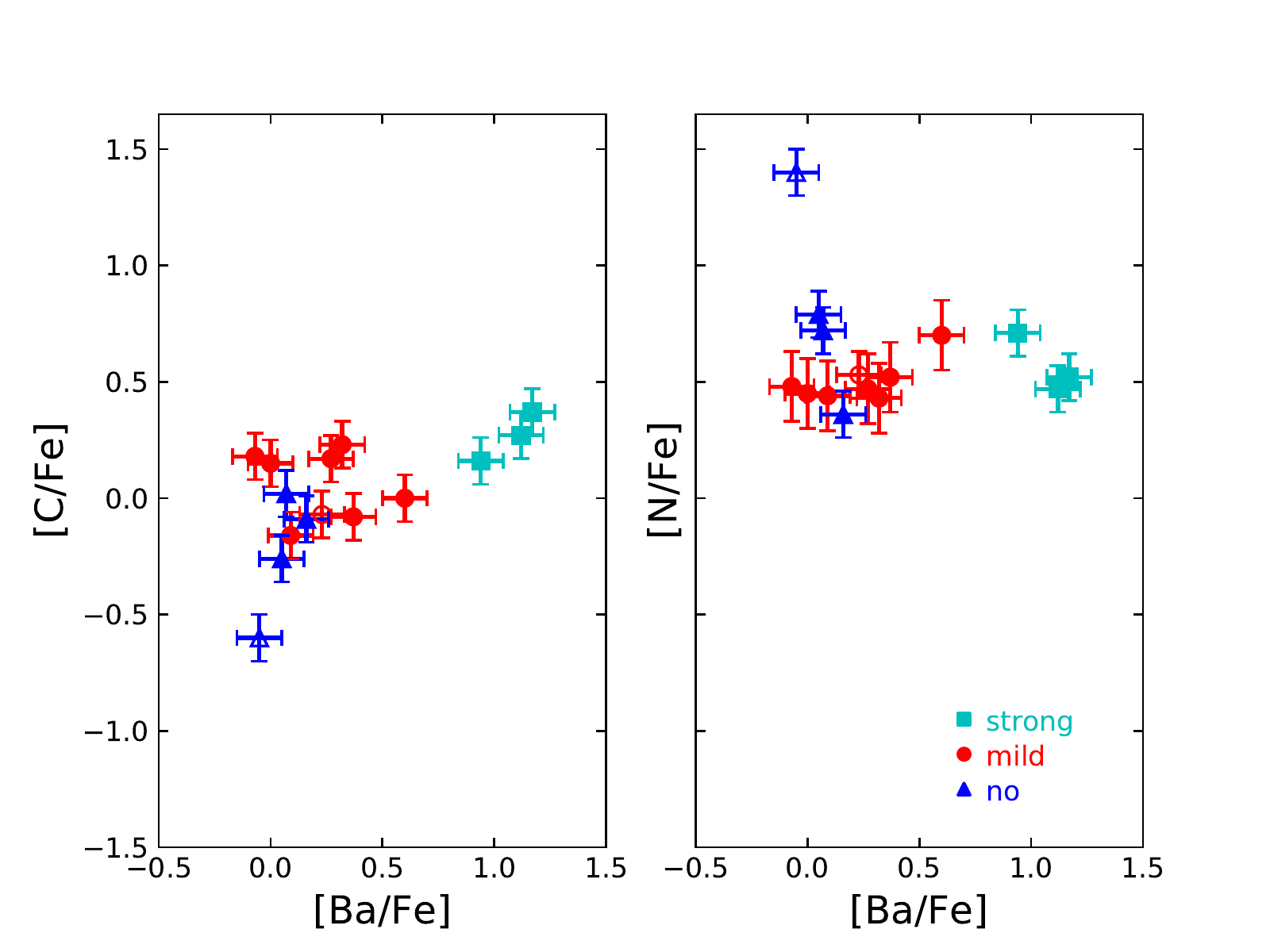}
\includegraphics[width=9cm]{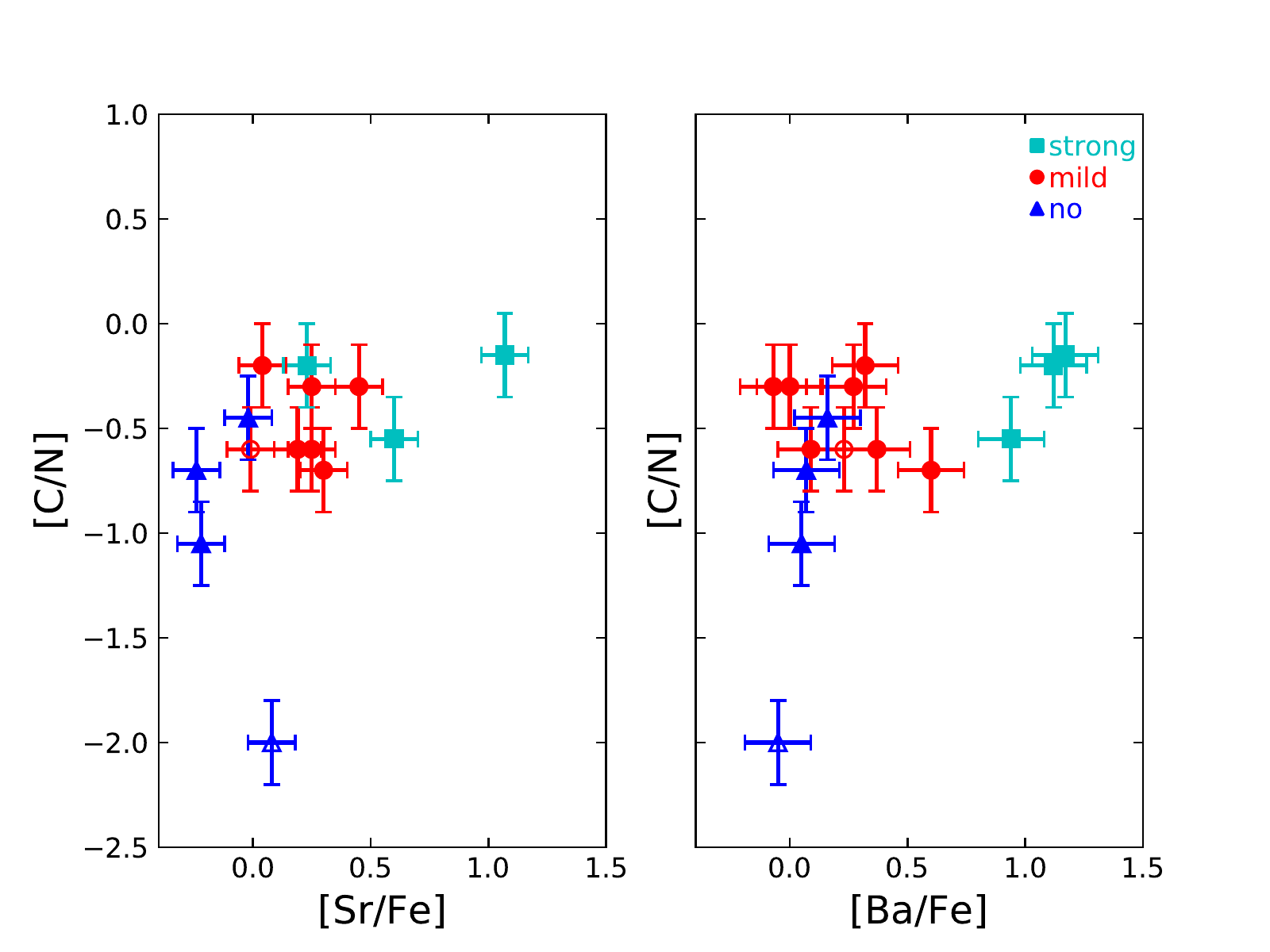}
\caption{Various diagnostics involving C and N.}
\label{Fig:CNBa}
\end{figure}

As explained in Sect.~\ref{Sect:light-s}, 'no-s' stars show comparatively large enhancements in N and hence low [C/N] values (Fig.~\ref{Fig:CNBa}). The low [C/N] values in these stars may be the result of an efficient mixing as they ascend the first giant branch (RGB), since their [N/Fe] ratio increases with luminosity along the RGB.

\section{Automatic vs. manual classification}
\label{Sect:TheCannon}

Table~\ref{Tab:programme_stars} reveals that 4 stars out of 15 have in fact been erroneously flagged as mild barium stars by the machine-learning method used by 
N19. 
The topic of this section is to identify what may be learned about the power of such machine-learning methods: can we identify why machine-learning led to such a large fraction of "false positives"? What are the properties of these false positives, with respect to either the accuracy of the atmospheric parameters delivered by {\it "The Cannon"} or the properties of the Ba and Sr lines used (are they saturated or not?)? 

Table~\ref{Tab:programme_stars} compares the atmospheric parameters obtained by N19 and by our high-resolution abundance study. Effective temperatures and gravities are generally in good agreement (the worst discrepancy is 200~K for $T_{\rm eff}$ in HIP 69788 and TYC 2250--1047--1, and 1~dex for the gravity of HIP~69788). The metallicities are more discrepant, up to 0.3~dex (HIP 69788, BD +44$^\circ$575, and TYC 4837$-$925$-$1). 
However, these discrepant atmospheric parameters are not restricted to those cases where we found no s-process enrichments. Therefore, we believe that false positives do not result from possible inaccuracies in the machine-learning atmospheric parameters.

N19 used strong \ion{Sr}{II} and \ion{Ba}{II} lines to derive Sr and Ba abundances. From the \ion{Sr}{II} lines at 4077 and 4215~\AA~, N19 found [Sr/Fe] values in the range 0.8 -- 1.0~dex for the four stars with no s-process enrichment in our analysis. 
Figure~\ref{Fig:Sr-line} reveals that the \ion{Sr}{II} line at 4215~\AA~ used by 
N19 is not only saturated, but also blended by the CN band with its band head at 4216~\AA. When this band is strong (i.e., in K giants with a large N abundance [N/Fe]~$\ge 0.7$), it likely causes false "Sr-only" positives. 

\section{Conclusions}
\label{Sect:conclusions}

A detailed abundance analysis of fifteen suspected Ba stars from N19 has been carried out. It was found that three of them are strongly enhanced with s-process elements, eight are mildly enhanced and the remaining four show no enhancement in s-process elements. The machine-learning technique used earlier by N19 on low-resolution LAMOST spectra classified thirteen among these fifteen stars as Sr-only candidates. Instead, our traditional approach based on an individual spectroscopic analysis of high-resolution spectra revealed that four of these 13 stars do not have significant overabundances of any s-process elements.  We investigated the possible reasons for the high incidence of the Sr-abundance tag obtained by the machine-learning technique. We found that the Sr lines used by N19 are generally saturated, thus leading to spurious overabundances. Neither has the possibility of a nitrogen enhancement in these stars, as revealed by our analysis, been considered by the N19 analysis.  Because of the blend between the \ion{Sr}{II}
4215.5~\AA~  line and the CN band head, a high N abundance, if overlooked, may lead to spurious Sr overabundances. 

We spectroscopically identified two strong  Ba dwarfs in the sample, further confirmed by their location in the HR diagram. We found significant radial velocity variations in five objects, two in the strong Ba class, one in the mild Ba class, and more surprisingly, two in the no-s class.
All the sample stars have Galactic thin-disk kinematic signatures, as evident from their location in the Toomre diagram.  

We also compared the properties of mild and strong Ba stars. Various heavy-s abundances revealed a sensitivity to metallicity, since all strong Ba stars have sub-solar metallicities. Carbon and N abundances seem to behave differently in the two groups, with non-Ba stars having a tendency to be C-poor but N-rich.  

\begin{acknowledgements}
 D.K. acknowledges the financial support from CSIR-India through file No.13(9086-A)2019-Pool. SVE thanks the Fondation ULB for its support. 
 The {\it Mercator} telescope is operated thanks to grant number G.0C31.13 of the FWO under the “Big Science” initiative of the Flemish governement. Based on observations obtained with the HERMES spectrograph, supported by the Fund for Scientific Research of Flanders (FWO), the Research Council of K.U.Leuven, the Fonds National de la Recherche Scientifique (F.R.S.- FNRS), Belgium, the Royal Observatory of Belgium, the Observatoire de Genève, Switzerland and the Thüringer Landessternwarte Tautenburg, Germany. This research has made use of the SIMBAD database, operated at CDS, Strasbourg, France and NASA ADS, USA. LS and SG are senior research associates from F.R.S.- FNRS (Belgium).

\end{acknowledgements}

\bibliographystyle{aa}
\bibliography{references}

\begin{appendix}

\section{Abundances}
The following tables present the elemental abundances of the programme stars. NLTE abundance corrections are applied when available, i.e. for O and Sr. We used the same atomic and molecular lines presented in \citet{Karinkuzhi2018,Karinkuzhi2021}. 
%The solar abundances are taken from \citet{Asplund2009}.

\begin{table*}[h!]
\caption{Light element abundances.
\label{Tab:light_abundances}}
\begin{tabular}{lrrrrrrrrr}
\hline
\\
Star Name   &[Ca/Fe] &[Sc/Fe]  &[Ti/Fe] & [V/Fe]&[Cr/Fe] &[Ni/Fe]&[Cu/Fe] & [Zn/Fe]\\
\hline
\\
HD 7863           &  0.03 &   -- &   0.12  & 0.44 & 0.03   &$-$0.30   &  $-$0.22  & $-$0.49   \\
HIP 69788         & 0.30&--0.01&0.09&--0.26&0.40&$-$0.03&0.15&0.03 \\
TYC 3144$-$1906$-$1   & $-$0.21&-0.02&--0.12&0.08&0.09&$-$0.24&$-$0.06&$-$0.43 \\
TYC 4684$-$2242$-$1   &0.26&0.05&0.25& 0.47& 0.21&$-$0.07& 0.11& $-$0.11 \\
BD$-$07 402         &0.07&0.11&0.16&0.38&0.11&$-$0.01&0.07&0.05 \\
BD+44 575         &0.41&0.15&0.50&--&0.41&0.01& 0.01&$-$0.31  \\
TYC 22$-$155$-$1      &0.36&--& 0.40& 0.47&0.16&0.13&0.21&0.24\\
TYC 2913$-$1375$-$1   &0.27&0.06&0.01&0.06&0.57&0.39&0.11&0.08 \\
TYC 3305$-$571$-$1    &0.31&0.20&0.10&0.05&0.10&$-$0.17&0.21&0.09\\
TYC 752$-$1944$-$1    &0.04&--& $-$0.17&0.05&0.14& 0.01&-0.11&0.02 \\
TYC 4837$-$925$-$1    &0.03&0.22&0.17&0.19&0.28& 0.05&$-$0.02&-0.04\\
TYC 3423$-$696$-$1    & 0.24& --&0.23&0.45&--& $-$0.24& 0.09& $-$0.03\\
TYC 2250$-$1047$-$1   & 0.21&0.20&0.20&0.30&0.31&0.03&$-$0.24&0.19\\
TYC 2955$-$408$-$1    &0.35&--& 0.24&--&0.35&--&0.20&0.43  \\
TYC 591$-$1090$-$1    & 0.21&0.15&0.20&0.10&0.36&$-$0.22&0.11&$-$0.26\\

\hline
\end{tabular}
%\end{center}
\end{table*}

\setcounter{table}{1}

{\footnotesize
\begin{table*}\small
\caption{Elemental abundances
\label{Tab:abundances}}
\begin{tabular}{lccccrccrrcrcrr}    
\hline
\\
\multicolumn{2}{c}{}& \multicolumn{4}{c}{BD $-07^\circ 402$} &&\multicolumn{3}{c}{BD $+44^\circ 575$}&& \multicolumn{3}{c}{HD 7863} \\
\cline{4-6}\cline{8-10}\cline{12-14}\\
 &    Z  &    log$_{\odot}{\epsilon}^a$ & log${\epsilon}$&$\sigma_{l}$ (N)& [X/Fe] $\pm~\sigma_{t}$ & & log${\epsilon}$&$\sigma_{l}$(N)& [X/Fe]~$\pm~ \sigma_{t}$ &&  log${\epsilon}$&$\sigma_{l}$(N)& [X/Fe]~$\pm~ \sigma_{t}$ \\
 &       &                         &                  &        &        \\

\hline
Li & 3  & 1.05   &  1.30 & 0.1(1) &  0.36 $\pm$ 0.18&&  --  &  --     &   --   &&  --  &  --     &   --  & \\
C  & 6  & 8.43   &  8.25 & 0.1(4) & $-$0.07 $\pm$ 0.11 &&   8.15 &  0.10(4)& 0.17 $\pm$ 0.11   &&8.10 & 0.10(1) & $-$0.26 $\pm$ 0.13 &  \\
$^{12}$C/$^{13}$C&  --   & --     &  --   &  --& 19 && -- &-- &  13     &&--  &  --     &   19  &  \\
N  & 7  & 7.83   & 8.25  & 0.1(10)&  0.53 $\pm$ 0.06 &&  7.85  & 0.05(15)& 0.47 $\pm$ 0.05  && 8.55 & 0.02(12)& 0.79 $\pm$ 0.05 &  \\
O  & 8  & 8.69   & 8.8   & 0.1(1) &  0.22 $\pm$ 0.12 &&  8.60  & 0.10(1) & 0.36 $\pm$ 0.13  && 8.65 & 0.10(1) & 0.03 $\pm$ 0.12 &  \\
Na & 11 & 6.24   & 6.53  & 0.08(2)&  0.40 $\pm$ 0.12  &&  6.38  & 0.08(2) & 0.59 $\pm$ 0.12 && 6.44 & 0.11(4) & 0.27 $\pm$ 0.12  &  \\
Mg & 12 & 7.60   &  --   &   --   &  --   &&  8.10: & 0.10(2) & 0.90 $\pm$ 0.15  && 7.44 & 0.10(3) &$-$0.09 $\pm$ 0.15 &  \\
Rb & 37 & 2.52   & 2.50  &  0.10(2)& 0.09 $\pm$ 0.16 &&--   & --      & --     &&2.30 & 0.10(2) & $-$0.15  $\pm$ 0.16&  \\
Sr & 38 & 2.87& 2.75  &   0.16(3)& $-$0.01 $\pm$ 0.13 &&  2.67  & 0.00(2) & 0.25 $\pm$ 0.11   &&2.58 & 0.10(3) & $-$0.22 $\pm$ 0.10 &  \\
Y  & 39 & 2.21   & 2.09  & 0.14(7)&$-$0.01 $\pm$ 0.14&&  1.72  & 0.04(6) & $-$0.04 $\pm$ 0.13&& 1.76 & 0.20(9)& $-$0.38 $\pm$ 0.14 &  \\
Zr & 40 & 2.58   & 2.75  & 0.07(3)& 0.28 $\pm$ 0.13  &&  2.61  & 0.04(4) & 0.48 $\pm$ 0.12   &&2.50 & 0.10(2) & $-$0.15 $\pm$ 0.14 &  \\
Nb & 41 & 1.46   & --     & --      & --     &&  1.48  & 0.12(3) & 0.47 $\pm$ 0.15   && --   & --       &   -   &  \\
Ba & 56 & 2.18   & 2.30  & 0.1(2) & 0.23 $\pm$ 0.07  &&  2.00  & 0.10(1) & 0.27 $\pm$ 0.10  &&2.15 & 0.15(4) &  0.05 $\pm$ 0.08 &  \\
La & 57 & 1.10   & 1.30  & 0.10(6)& 0.31 $\pm$ 0.10  &&  1.14  & 0.05(8) & 0.49 $\pm$ 0.09   &&0.96 & 0.06(4) & $-$0.07 $\pm$ 0.10 &  \\
Ce & 58 & 1.58   & 1.64  & 0.08(6)& 0.17 $\pm$ 0.07 &&  1.50  & 0.10(4) & 0.37 $\pm$ 0.08   &&1.34 & 0.12(7) & $-$0.14 $\pm$ 0.07 &  \\
\hline
\\
\multicolumn{2}{c}{}& \multicolumn{4}{c}{HIP 69788} && \multicolumn{3}{c}{TYC 22$-$155$-$1} && \multicolumn{3}{c}{TYC 2250$-$1047$-$1} \\
\cline{4-6}\cline{8-10}\cline{12-14}\\
 &    Z  &    log$_{\odot}{\epsilon}^a$ & log${\epsilon}$&$\sigma_{l}$(N)&  [X/Fe]~$\pm~ \sigma_{t}$ && log${\epsilon}$&$\sigma_{l}$(N)& [X/Fe]~$\pm~ \sigma_{t}$ && log${\epsilon}$&$\sigma_{l}$(N)& [X/Fe]~$\pm~ \sigma_{t}$ \\
 &       &                         &                  &        &        \\
\hline
C  & 6  & 8.43 & 8.30 & 0.10(4) &$-$0.09 $\pm$ 0.11 & & 8.45  & 0.10(4) &  0.23 $\pm$ 0.11 &&  8.25 & 0.10(4) & 0.37 $\pm$ 0.11 &  \\
N  & 7  & 7.83 & 8.15 & 0.05(15)&0.36 $\pm$ 0.05   & & 8.05  & 0.10(9)& 0.43 $\pm$  0.06  &&  7.80 & 0.10(10)& 0.52 $\pm$ 0.06 &  \\
O  & 8  & 8.69 & 8.90 & 0.10(2) &0.25 $\pm$ 0.10   & & 8.90  & 0.10(1) & 0.44 $\pm$ 0.13   &&  8.60 & 0.10(1) & 0.46 $\pm$  0.13&  \\
Na & 11 & 6.24 & 6.30 & 0.10(2) &0.10 $\pm$ 0.13   & & 6.35  & 0.05(2) & 0.31 $\pm$ 0.11  &&  5.85 & 0.10(2) & 0.16 $\pm$ 0.13 &  \\
Mg & 12 & 7.60 &  --  &  --     &   --   & & 8.60: & 0.10(1) & 1.20 $\pm$ 0.15  &&   --   &--    & --   &  \\
Rb & 37 & 2.52 & 2.50 & 0.10(2) &0.02 $\pm$ 0.16   & &--     &  --     & --    &&  2.10 & 0.10(1) & 0.13   $\pm$ 0.17& \\
Sr & 38 & 2.87& 2.81  &   0.16(2)& $-$0.02 $\pm$ 0.15  &&  2.71  & 0.19(3) & 0.04 $\pm$ 0.14   &&3.39 & 0.19(2) & 1.07 $\pm$ 0.16 &  \\
Y  & 39 & 2.21 & 1.79 & 0.09(5) &$-$0.38 $\pm$ 0.13 & & 1.78  & 0.06(4) &$-$0.23 $\pm$ 0.13&&  2.10 & 0.13(5) & 0.44 $\pm$ 0.14 &  \\
Zr & 40 & 2.58 & 2.20 & 0.10(3) &$-$0.34 $\pm$ 0.13  & & 2.55  & 0.05(4) &0.17  $\pm$ 0.12  &&  2.80 & 0.10(2) & 0.77 $\pm$ 0.14  &  \\
Ba & 56 & 2.18 & 2.30 & 0.10(1) &0.16 $\pm$ 0.10    & & 2.30  & 0.10(2) &0.32 $\pm$ 0.07  &&  2.80 & 0.10(3) & 1.17 $\pm$  0.07 & \\
La & 57 & 1.10 & 1.00 & 0.10(4) &$-$0.06 $\pm$ 0.11 & & 1.21  & 0.12(9) &0.31 $\pm$ 0.10  &&  1.73 & 0.11(8) & 1.18 $\pm$ 0.10 &  \\
Ce & 58 & 1.58 & 1.30 & 0.10(2) &$-$0.24 $\pm$ 0.09 & & 1.50  & 0.10(2) &0.12 $\pm$ 0.09  &&  2.00 & 0.07(4) & 0.97 $\pm$ 0.07 &  \\
\hline
\end{tabular}

$^{a}$ Asplund et al. (2009) \\
$:$ Uncertain abundances due to noisy/blended region\\
\end{table*}
}

\setcounter{table}{1}
{\footnotesize
\begin{table*}\small
\caption{Elemental abundances
}
\begin{tabular}{lccccrccrrcrcrr}  
\hline
\\
\multicolumn{2}{c}{}& \multicolumn{4}{c}{TYC 2913$-$1375$-$1} &&\multicolumn{3}{c}{TYC 2955$-$408$-$1}&& \multicolumn{3}{c}{TYC 3144$-$1906$-$1} \\
\cline{4-6}\cline{8-10}\cline{12-14}\\
 &    Z  &    log$_{\odot}{\epsilon}^a$ & log${\epsilon}$&$\sigma_{l}$ (N)& [X/Fe] $\pm~\sigma_{t}$ & & log${\epsilon}$&$\sigma_{l}$(N)& [X/Fe]~$\pm~ \sigma_{t}$ &&  log${\epsilon}$&$\sigma_{l}$(N)& [X/Fe]~$\pm~ \sigma_{t}$ \\
 &       &                         &                  &        &        \\
\hline
Li & 3  & 1.05   &  -- & -- &  -- &&  --  &  --     &   --   &&  0.60  &  0.10(1) & $-$0.32 $\pm$ 0.18  & \\
C  & 6  & 8.43   & 8.00 & 0.10(3) & 0.18 $\pm$ 0.11 & &  8.2 & 0.10(4) & 0.16 $\pm$ 0.11   && 7.70&0.10(1)&$-$0.60 $\pm$ 0.14 &   \\
N  & 7  & 7.83   & 7.70 & 0.10(10)& 0.48 $\pm$ 0.06 & & 8.15 & 0.06(15)& 0.71 $\pm$ 0.05   && 9.10&    & 1.40 $\pm$ 0.11    &   \\
O  & 8  & 8.69   &  --  &  --     &   -- & & 8.60 & 0.10(1) & 0.33 $\pm$ 0.13   && --  & -- &   --&   \\
Na & 11 & 6.24   & 8.60 & 0.10(1) & 0.52 $\pm$ 0.15 & & 6.15 & 0.10(2) & 0.30 $\pm$  0.13  && 6.60 & 0.10(2) & 0.49 $\pm$ 0.13&    \\
Mg & 12 & 7.60   &  --  & --      &      & & 8.20 & 0.10(3) & 0.99 $\pm$ 0.15   && --   &  --  &  --   &   \\
Rb & 37 & 2.52   &  --  & --      &--     & & 2.48 & 0.08(2) & 0.35 $\pm$ 0.15 && 2.48 & 0.08(2) &0.09 $\pm$ 0.16     &   \\
Sr & 38& 2.87& -- & -- & -- & & 3.08 & 0.19(2) & 0.60 $\pm$ 0.16   && 2.82 & 0.05(2) & 0.08 $\pm$ 0.15    &   \\
Y  & 39 & 2.21   & 1.25 & 0.21(5) &$-$0.35 $\pm$ 0.16 & & 2.43 & 0.13(9)& 0.61 $\pm$ 0.14 && 1.88 & 0.25(5) &$-$0.20 $\pm$ 0.17 &   \\
Zr & 40 & 2.58   & 2.28 & 0.04(4) & 0.31 $\pm$ 0.12 & & 3.15 & 0.12(3) & 0.96 $\pm$ 0.14  && 2.67 & 0.11(6) & 0.22 $\pm$ 0.13   &   \\
Nb & 41 & 1.46   &--	&--     & --   & & 1.98 & 0.06(4) & 0.95 $\pm$ 0.15   &&  --   & --       & -- &   \\
Ba & 56 & 2.18   & 1.50 & 0.14(2) &$-$0.07 $\pm$ 0.10& & 2.73 & 0.05(3) & 0.94 $\pm$ 0.05  && 2.00 &0.10(2) & $-$0.05 $\pm$ 0.07 &   \\
La & 57 & 1.10   & 0.85 & 0.14(7) & 0.36 $\pm$ 0.11 & & 1.45 & 0.07(9) & 0.74 $\pm$ 0.10  && 1.16 & 0.16(10)& 0.19 $\pm$ 0.11    &   \\
Ce & 58 & 1.58   & 1.12 & 0.11(5) & 0.15 $\pm$ 0.07 & & 1.84 & 0.09(8) & 0.65 $\pm$ 0.07  && 1.53 & 0.05(4) & 0.08 $\pm$ 0.06    &   \\
\hline
\\
\multicolumn{2}{c}{}& \multicolumn{4}{c}{TYC 3305$-$571$-$1} && \multicolumn{3}{c}{TYC 3423$-$6966$-$1} && \multicolumn{3}{c}{TYC 4684$-$2242$-$1} \\
\cline{4-6}\cline{8-10}\cline{12-14}\\
 &    Z  &    log$_{\odot}{\epsilon}^a$ & log${\epsilon}$&$\sigma_{l}$(N)&  [X/Fe]~$\pm~ \sigma_{t}$ && log${\epsilon}$&$\sigma_{l}$(N)& [X/Fe]~$\pm~ \sigma_{t}$ && log${\epsilon}$&$\sigma_{l}$(N)& [X/Fe]~$\pm~ \sigma_{t}$ \\
 &       &                         &                  &        &        \\
\hline
C  & 6  & 8.43 &  8.30& 0.10(4)  &$-$0.08 $\pm$ 0.11 && 8.60  & 0.10(3) & 0.15 $\pm$ 0.12 && 8.40 & 0.10(4) & 0.02 $\pm$ 0.11  &   \\
N  & 7  & 7.83 & 8.30 & 0.10(10) & 0.52 $\pm$ 0.07  && 8.30  & 0.15(15)& 0.45 $\pm$ 0.06&& 8.50 & 0.06(15)& 0.72 $\pm$ 0.05  &   \\
O  & 8  & 8.69 & 8.90 & 0.10(1)  & 0.29 $\pm$ 0.13  && 8.90  & 0.10(1) & 0.22 $\pm$ 0.12 && 8.90 & 0.10(1) & 0.26 $\pm$ 0.12  &   \\
Na & 11 & 6.24 & 6.53 & 0.08(2)  & 0.34 $\pm$ 0.12  && 6.60  & 0.10(2) & 0.34 $\pm$ 0.13 && 6.60 & 0.10(2) & 0.41 $\pm$ 0.13 &   \\
Mg & 12 & 7.60 & --    &--       & --     &&--     & --      &--    && 8.20:& 0.10(1) & 0.65 $\pm$ 0.15  &   \\
Rb & 37 & 2.52 & 2.55 & 0.10(2)  & 0.08 $\pm$ 0.16  && 2.80  & 0.10(2) & 0.26 $\pm$ 0.16 && 2.60 & 0.10(1) & 0.13 $\pm$ 0.17   &   \\
Sr & 38 & 2.87& 3.05 & 0.10(2)& 0.25 $\pm$ 0.12  && 3.34  & 0.06(2) & 0.45 $\pm$ 0.10 && 2.58 & 0.28(2) & $-$0.24 $\pm$ 0.22  &   \\
Y  & 39 & 2.21 & 2.18 & 0.14(8)  & 0.02 $\pm$ 0.14  && 2.36  & 0.19(7) & 0.13 $\pm$ 0.15 && 2.04 & 0.14(8) &$-$0.12 $\pm$ 0.14&   \\
Zr & 40 & 2.58 & 2.78 & 0.02(3)  & 0.25 $\pm$ 0.12  && 2.80  & 0.10(3) & 0.20 $\pm$ 0.14 && 2.75 & 0.05(4) & 0.22 $\pm$ 0.12  &   \\
Nb & 41 &1.46  &   -- &  --      & --     && --        &   --   & &&1.80:& 0.10(1) & 0.39 $\pm$ 0.15   &   \\
Ba & 56 & 2.18 & 2.50 & 0.10(2)  & 0.37 $\pm$ 0.08  && 2.20  & 0.10(1) & 0.00 $\pm$ 0.11&& 2.20 & 0.10(2) & 0.07 $\pm$ 0.07  &   \\
La & 57 & 1.10 & 1.53 & 0.04(8)  & 0.48 $\pm$ 0.10   && 1.27  & 0.12(5) & 0.15 $\pm$ 0.11 && 1.26 & 0.04(5) & 0.21 $\pm$ 0.10 &   \\
Ce & 58 & 1.58 & 1.88 & 0.15(5)  & 0.35 $\pm$ 0.10  && 1.77  & 0.05(3) & 0.02 $\pm$ 0.06 && 1.67 & 0.11(6) & 0.14 $\pm$ 0.07 &   \\
\hline
\\
\multicolumn{2}{c}{}& \multicolumn{4}{c}{TYC 4837$-$925$-$1} &&\multicolumn{3}{c}{TYC 591$-$1090$-$1}&& \multicolumn{3}{c}{TYC 752$-$1944$-$1} \\
\cline{4-6}\cline{8-10}\cline{12-14}\\
 &    Z  &    log$_{\odot}{\epsilon}^a$ & log${\epsilon}$&$\sigma_{l}$ (N)& [X/Fe] $\pm~\sigma_{t}$ & & log${\epsilon}$&$\sigma_{l}$(N)& [X/Fe]~$\pm~ \sigma_{t}$ &&  log${\epsilon}$&$\sigma_{l}$(N)& [X/Fe]~$\pm~ \sigma_{t}$ \\
 &       &                         &                  &        &        \\
\hline
C  & 6  & 8.43 & 8.00 & 0.10(4) &$-$0.16 $\pm$ 0.11 && 8.40 & 0.10(4) & 0.27 $\pm$ 0.11  &&  8.35 & 0.05(4)  & 0.00 $\pm$ 0.10   & \\
N  & 7  & 7.83 & 8.00 & 0.10(15)& 0.44 $\pm$ 0.06  && 8.15 & 0.15(15)& 0.47 $\pm$ 0.07  &&  8.45 & 0.05(15) & 0.70 $\pm$ 0.05   &\\
O  & 8  & 8.69 & 8.60 & 0.10(1) & 0.18 $\pm$ 0.12  && 8.40 & 0.20(2) & 0.01 $\pm$ 0.16  &&  8.75 & 0.10(1)  & 0.17 $\pm$ 0.12    &\\
Na & 11 & 6.24 & 6.25 & 0.05(2) & 0.28 $\pm$ 0.11  && 6.05 & 0.05(2) & 0.11 $\pm$ 0.11  &&  6.45 & 0.10(2)  & 0.28 $\pm$  0.13   & \\
Mg & 12 & 7.60 & --   &  --     &--     && --   & --      & --    &&  7.75 & 0.15(2)  & 0.23 $\pm$ 0.15    &\\
Rb & 37 & 2.52   & --   &  --     & --    && --   & --      &--     &&  2.80 & 0.10(1)  & 0.28 $\pm$ 0.17 & \\
Sr & 38 & 2.87 & 2.79 & 0.18(2) &0.19 $\pm$ 0.16   && 2.80 & 0.10(1) & 0.23 $\pm$ 0.13  &&  3.09 & 0.18(1)  & 0.30 $\pm$ 0.12& \\
Y  & 39 & 2.21 & 1.74 & 0.12(6) &$-$0.20 $\pm$ 0.14&& 3.03 & 0.04(9)& 1.12 $\pm$ 0.13  &&  2.60 & 0.10(5)  & 0.47 $\pm$ 0.14    & \\
Zr & 40 & 2.58 & 2.48 & 0.02(3) &0.17 $\pm$ 0.16   && 3.37 & 0.07(5) & 1.09 $\pm$ 0.12  &&  3.09 & 0.02(4)  & 0.59 $\pm$ 0.12    & \\
Ba & 56 & 2.18 & 2.00 & 0.10(2) &0.09 $\pm$ 0.07   && 3.00 & 0.15(2) & 1.12 $\pm$ 0.11  &&  2.70 & 0.10(1)  & 0.60 $\pm$ 0.10    & \\
La & 57 & 1.10 & 1.08 & 0.04(6) &0.25 $\pm$ 0.09   && 1.95 & 0.14(8) & 1.15 $\pm$ 0.11  &&  1.59 & 0.11(5)  & 0.57 $\pm$ 0.11   & \\
Ce & 58 & 1.58 & 1.50 & 0.10(5) & 0.19 $\pm$ 0.07  && 2.36 & 0.16(4) & 1.08 $\pm$ 0.10  &&  2.03 & 0.06(7)  & 0.53 $\pm$ 0.06   & \\
\hline
\end{tabular}

$^{a}$ Asplund et al. (2009) \\
$:$ Uncertain abundances due to noisy/blended region\\
\end{table*}
}

\end{appendix}
\end{document}